\documentclass[12pt,preprint]{aastex}
\usepackage[numberedappendix]{emulateapj5}

\newcommand{\co}          {{\rm $^{12}$CO}}
\newcommand{\hi}          {\mbox{\rm \ion{H}{1}}}
\newcommand{\hii}         {\mbox{\rm \ion{H}{2}}}
\newcommand{\htwo}        {H$_{2}$}
\newcommand{\jone}        {($J=1\rightarrow0$)}
\newcommand{\percmsq}     {cm$^{-2}$}
\newcommand{\kms}         {km~s$^{-1}$}
\newcommand{\jykms}       {Jy~km~s$^{-1}$}
\newcommand{\gpercmcu}    {g~cm$^{-3}$}
\newcommand{\ha}          {\mbox{H$\alpha$}}
\newcommand{\mlk}         {\mbox{$M_{\odot}/L_{\odot K}$}}
\newcommand{\mlr}         {\mbox{$M_{\odot}/L_{\odot R}$}}
\newcommand{\mslk}        {\mbox{$M_{*}/L_{K}$}}
\newcommand{\msl}         {\mbox{$M_{*}/L$}}
\newcommand{\mslr}        {\mbox{$M_{*}/L_{R}$}}
\newcommand{\hr}          {\mbox{$^h$}}
\renewcommand{\min}       {\mbox{$^m$}}
\renewcommand{\sec}       {\mbox{$^s$}}
\newcommand{\ad}          {($\alpha,\delta$)}
\newcommand{\surfb}       {mag~arcsecond$^{-2}$}
\newcommand{\vco}         {$v_{\mbox{{\tiny CO}}}$}
\newcommand{\vha}         {$v_{\mbox{{\tiny \ha}}}$}
\newcommand{\mpc}         {$M_{\odot}$~pc$^{-3}$}
\newcommand{\vhirot}      {v_{\mbox{\tiny HI},rot}}

\newcommand{\vmolrot}     {v_{\mbox{\tiny H$_{2}$},rot}}
\newcommand{\rotcur}      {{\sc rotcur}}
\newcommand{\rotcurshape} {{\sc rotcurshape}}
\newcommand{\ringfit}     {{\sc ringfit}}
\newcommand{\alphadm}     {\alpha_{\mbox{{\tiny DM}}}}
\newcommand{\alphatot}    {\alpha_{\mbox{{\tiny TOT}}}}
\newcommand{\rhodm}       {\rho_{\mbox{{\tiny DM}}}}
\newcommand{\rhotot}      {\rho_{\mbox{{\tiny TOT}}}}

\slugcomment{Accepted for publication in \emph{The Astrophysical
Journal}} 
\shorttitle{Dark Matter Halo of NGC~2976}
\shortauthors{Simon et al.}

\begin{document}

\title{High-Resolution Measurements of the Dark Matter Halo of
NGC~2976: Evidence for a Shallow Density Profile\altaffilmark{1}}

\author{Joshua D. Simon\altaffilmark{2}, Alberto
D. Bolatto\altaffilmark{2}, Adam Leroy\altaffilmark{2}, and Leo Blitz}

\altaffiltext{1}{Based on observations carried out at the WIYN
Observatory.  The WIYN Observatory is a joint facility of the
University of Wisconsin-Madison, Indiana University, Yale University,
and the National Optical Astronomy Observatory.}

\altaffiltext{2}{Visiting Astronomer, Kitt Peak National Observatory,
National Optical Astronomy Observatory, which is operated by the
Association of Universities for Research in Astronomy, Inc. (AURA)
under cooperative agreement with the National Science Foundation.  }

\affil{Department of Astronomy, University of California at Berkeley}
\affil{601 Campbell Hall, CA  94720}
\email{jsimon@astro.berkeley.edu, bolatto@astro.berkeley.edu,
aleroy@astro.berkeley.edu, blitz@astro.berkeley.edu}

\begin{abstract}
We have obtained two-dimensional velocity fields of the dwarf spiral
galaxy NGC~2976 in \ha\ and CO.  The high spatial ($\sim 75$ pc) and
spectral (13 \kms\ and 2 \kms, respectively) resolution of these
observations, along with our multicolor optical and near-infrared
imaging, allow us to measure the shape of the density profile of the
dark matter halo with good precision.  We find that the total
(baryonic plus dark matter) mass distribution of NGC~2976 follows a
$\rhotot \propto r^{-0.27 \pm 0.09}$ power law out to a radius of 1.8
kpc, assuming that the observed radial motions provide no support.
The density profile attributed to the dark halo is even shallower,
consistent with a nearly constant density of dark matter over the
entire observed region.  A maximal disk fit yields an upper limit to
the K-band stellar mass-to-light ratio (\mslk) of
$0.09^{+0.15}_{-0.08} \mlk$ (including systematic uncertainties), with
the caveat that for $\mslk > 0.19 \mlk$ the dark matter density
increases with radius, which is unphysical.  Assuming $0.10 \mlk
\lesssim \mslk \le 0.19 \mlk$, the dark matter density profile lies
between $\rhodm \propto r^{-0.17}$ and $\rhodm \propto r^{-0.01}$.
Therefore, independent of any assumptions about the stellar disk or
the functional form of the density profile, NGC~2976 does not contain
a cuspy dark matter halo.  We also investigate some of the systematic
effects that can hamper rotation curve studies, and show that 1)
longslit rotation curves are far more vulnerable to systematic errors
than two-dimensional velocity fields, 2) NGC~2976 contains radial
motions that are as large as 90~\% of the rotational velocities at
small radii, and 3) the \ha\ and CO velocity fields of NGC~2976 agree
within their uncertainties, with a typical scatter between the two
velocities of 5.3 \kms\ at any position in the galaxy.
\end{abstract}


\section{INTRODUCTION}

The apparent disagreement between the observed dark matter density
profiles of dwarf and low-surface brightness (LSB) galaxies and the
density profiles predicted by numerical Cold Dark Matter (CDM)
simulations has been widely discussed by both theorists and observers
over the past several years \citep[e.g.,][]
{fp94,burkert95,nfw96,moore99b}.  However, there remains a disturbing
lack of consensus in the observational community on the actual shape
of the observed dark matter density profiles.  Many authors claim that
only constant-density cores are allowed by the observations
\citep{db01a,bs01,db01b,dbb02,swb02,wdbw03}.  On the other hand,
\citet{vdb00}, \citet{vdbs01}, and \citet[hereafter SMVB]{swaters02}
argue that most existing rotation curves are also consistent with
NFW-like ($\rho \propto r^{-1}$) central density cusps.  Even the very
highest resolution ($\lesssim 100$ pc) studies do not seem to be
converging on a single result; \citet{blo01} found $\rho \propto
r^{-0.3}$ in NGC~3109 and $\rho \propto r^{-0.5}$ in IC~2574 (ignoring
the stellar disk contributions to the rotation curves), \citet{us}
showed that NGC~4605 has a density profile $\rho \propto r^{-0.65}$,
and \citet*{wdbw03} determined that NGC~6822 contains an essentially
constant-density halo.

The recent study by SMVB shows that, in large part, the lack of
consensus among observers reflects ambiguities in the data themselves.
For the parameters of typical dwarf/LSB galaxy observations ($\sim 50$
\kms\ velocity resolution and $\sim 1$\arcsec\ seeing for longslit
\ha\ observations, and $\sim 2$ \kms\ velocity resolution and $\sim
15$\arcsec\ angular resolution for \hi\ interferometry), they find
that {\em most galaxies show central density profiles that are
consistent with any shape between $r^{0}$ and $r^{-1}$.}

We address this problem with a new study that combines a number of
techniques to overcome the observational challenges.  Our program
includes 1) two-dimensional velocity fields obtained at optical (\ha),
millimeter (CO), and centimeter (\hi) wavelengths, 2) high angular
resolution ($\sim5\arcsec$), 3) high spectral resolution ($\lesssim
10$ \kms), 4) multicolor optical and near-infrared photometry, and 5)
nearby dwarf galaxies as targets.  Observing completely independent
tracers of the velocity field at two or three different wavelengths
reduces our vulnerability to the systematic problems that can affect a
single tracer.  For example, \ha\ velocity fields can be distorted by
extinction, or by large-scale flows that are associated with star
formation, while existing \hi\ data generally suffer from beam
smearing.  Two-dimensional velocity fields also represent a major
improvement over the traditional longslit spectra, making the effect
of positioning errors negligible and allowing us to account for simple
noncircular motions.  High angular resolution is important because the
central cores described in the literature have typical radii of $\sim
1$ kpc, which corresponds to an angular size of $20.6 (\mbox{d}/10
\mbox{Mpc})^{-1}$ arcseconds.  In order to resolve this size scale and
minimize the impact of beam smearing on our conclusions, an angular
resolution element several times smaller is required.  High spectral
resolution is also beneficial because it results in more accurate
rotation curves.  Finally, our multicolor photometry plays a crucial
role in allowing us to attempt to realistically model the rotational
contribution from stellar disks instead of simply guessing an
appropriate mass-to-light ratio and assuming an exponential disk.

Target selection also has important effects on the strength of the
conclusions we will be able to draw.  We focus on very nearby objects
(D $<$ 10 Mpc) in order to maximize our physical resolution.  Dwarf
and LSB galaxies are the preferred targets for this type of study
because they are presumed to be the most dark-matter dominated
galaxies.  (Note that in this paper when we refer to dwarf galaxies,
we mean high-mass dwarf irregulars and low-mass spiral galaxies, not
dwarf spheroidals or ellipticals.)  LSB galaxies, though, tend to be
relatively distant and are necessarily quite faint, so they are
difficult to observe with sufficient resolution and sensitivity.
Dwarf galaxies, in comparison, are both bright and plentiful in the
nearby universe.  Dwarfs are traditionally presumed to be dark-matter
dominated at all radii \citep{cf88,cb89,jc90,mcr94}.  However, the
observations upon which this assumption is based were made at low
angular resolution.  Higher resolution observations of the inner
regions of dwarf galaxies show, as we discuss later, that stars can
dominate the kinematics of dwarf galaxies in their inner kpc
\citep[e.g.,][]{broeils92,swaters99,blo99,us}.  Comparable data for LSB
galaxies are scarce \citep[although see][]{smt00,swaters03}, but it is
possible that reliance on low-resolution observations
\citep[e.g.,][]{dbm97} could have caused an overstatement of the case
for dark matter domination in these galaxies as well.  Future studies
of LSB galaxies, featuring two-dimensional \ha\ spectroscopy and/or
$\le 100$ pc resolution \hi\ observations, if feasible, are desirable
both to investigate this question and to clarify the severity of the
cusp/core problem.

In a previous paper, we reported on a rotation curve study of the
dwarf spiral galaxy NGC~4605 \citep{us}.  In this paper, we present a
similar, but improved, study of a second nearby dwarf galaxy,
NGC~2976.  As before, we use high-resolution CO interferometry to
study the inner velocity field of the galaxy, but we have also
acquired high-resolution two-dimensional \ha\ data (instead of
longslit observations) to supplement the CO and extend the velocity
field out to larger radii.  In addition, we have obtained multicolor
optical imaging of this galaxy, which, combined with archival 2MASS
near-infrared images, enables us to accurately model the stellar
disk.

In the following section, we describe NGC~2976 and our observations
and data reduction.  In \S \ref{baryons}, we model the stellar and
gaseous disks.  In \S \ref{darkmatter}, we derive the rotation curve
of the galaxy and the density profile of its dark matter halo.  The
analysis routines that we use are presented in more detail in Appendix
\ref{algorithms}.  We discuss our results and their implications in \S
\ref{discussion}.  In \S \ref{systematics}, we describe some
systematic uncertainties that can affect rotation curve studies, and
test the robustness of our results against them.  We present our
conclusions in \S \ref{conclusions}.

\section{TARGET, OBSERVATIONS, AND DATA REDUCTION}
\label{observations}

\subsection{Properties of NGC~2976}

NGC~2976 is a regular Sc dwarf galaxy located in the M~81 group.
\citet{k02} measured a distance of $3.56 \pm 0.38$ Mpc using the Tip
of the Red Giant Branch (TRGB) method, and the Tully-Fisher distance
is $3.33 \pm 0.50$ Mpc (M. Pierce, private communication).  We adopt a
distance of 3.45 Mpc, which sets the conversion between physical and
angular scales to $16.7$ pc arcsec$^{-1}$.  NGC~2976 has absolute
magnitudes of M$_{B} = -17.0$ and M$_{K} = -20.2$, a heliocentric
velocity of $-0.8 \pm 1.8$ \kms, an inclination-corrected \hi\
linewidth W$_{20} = 165$ \kms, and a total mass of $3.5 \times
10^{9} M_{\odot}$, so it is somewhat less luminous and less massive
than the Large Magellanic Cloud.  The low systemic velocity is not a
problem for our observations because the galaxy is located at high
Galactic latitude, where there is little Milky Way CO emission, and no
Galactic \ha\ emission is visible.  In optical and near-infrared
images it is clear that NGC~2976 is a bulgeless, unbarred, pure disk
system (see Figure \ref{colorimage}), which makes it an ideal galaxy
for mass modeling.

\subsection{\ha\ Observations and Reductions}
\label{hadata}

Our \ha\ observations were obtained on the nights of 2002 March 20-21
at the 3.5 m WIYN telescope with the DensePak instrument.  DensePak is
an array of 94 2\farcs8-diameter fibers, fixed in a $30\arcsec \times
45\arcsec$ rectangle with a fiber-to-fiber spacing of 4\arcsec\ 
\citep{bsh98}.  Five of the fibers are broken and four are sky fibers,
placed at fixed positions outside the main array.  Thus, there are 85
data fibers covering almost the whole instrument footprint (see Figure
\ref{densepak}).  The fibers feed into the Bench Spectrograph, which
we used in its echelle mode to yield 13 \kms\ velocity resolution over
a 180 \AA\ range centered on \ha.  The detector was a $2048^{2}$ SITe
T2KC CCD.

We observed the galaxy at 13 positions to cover most of its optical
extent (see Figures \ref{haimage} and \ref{velfield}a).  The galaxy
was not visible on the guide camera at the telescope, so we acquired
the target by offsetting from a nearby bright star.  Each subsequent
position on the galaxy was observed by making a blind offset from the
previous position.  Integration times at each position were between 20
and 70 minutes, with just a single 20-minute exposure at most
positions.  Consecutive exposures at the same position were reduced
separately and then coadded.  Two of the fields were observed on both
nights, and one field was observed twice on the same night, but five
hours apart.  In these three cases, instead of assuming that the
positions observed were the same for the later observations as they
were for the earlier ones, we analyzed the frames entirely
independently.  We therefore had 16 observations of NGC~2976, yielding
a total of 1360 spectra, of which 1087 contained \ha\ emission at a
level of $3\sigma$ or higher.  Based on comparison with adjacent
fibers that contained brighter emission at similar velocities, we also
judged that 9 spectra containing emission at a significance level
between $2.2\sigma$ and $3\sigma$ represented real signal.  The median
detection level in the 1096 spectra that contained emission was
$27\sigma$ in integrated intensity, or $12\sigma$ at the peak of the
line.

The DensePak data were reduced in IRAF\footnote{IRAF is distributed by
the National Optical Astronomy Observatories, which is operated by the
Association of Universities for Research in Astronomy, Inc.  (AURA)
under cooperative agreement with the National Science Foundation.},
using the HYDRA package.  We subtracted a bias frame, removed cosmic
rays, interpolated over bad columns, and then extracted the spectra
with the task {\sc dohydra}.  The trace and response function for each
fiber and the relative transmission efficiencies were derived from a
set of flat field images, and wavelength calibration was provided by
spectra of a CuAr lamp.  Night-sky emission line wavelengths from
\citet{osterbrock96} and observations of a radial velocity standard
star were used to check the wavelength scale.  After extraction and
wavelength calibration, we averaged together the four sky fibers,
leaving out any sky spectra that were contaminated by emission lines
from the target galaxy.  We then removed a linear baseline, performed
a Gaussian fit to the averaged sky emission near \ha, and subtracted
the fit from all of the data fibers.  Some spectra contained
noticeable residuals at the wavelength of the sky \ha\ line after this
subtraction.  Sky residuals are easily distinguishable from real
signals because they are unresolved and always located in the same
four pixels.  If the residual overlapped with and was comparable in
strength to the \ha\ emission from NGC~2976, the spectrum was
discarded (29 spectra were thrown out because of this consideration).
This only occurred in places where the galaxy velocities were about
$-17$ \kms\ (see Figure \ref{velfield}).  Individual frames of the
same field were then averaged together (except for the cases noted
above), weighted by exposure time if it was clear, or signal-to-noise
ratio if there were clouds during the exposure.  Velocities were
calculated for each fiber by fitting a Gaussian to the observed \ha\
emission.  Typical linewidths are 34 \kms, and the median
uncertainties on the Gaussian fit centroids are 0.77 \kms; some fits
have uncertainties as small as 0.04 \kms, and a few are as large as 23
\kms.

It was obvious from comparing frames that were taken several hours
apart or on different nights that the telescope positioning accuracy
for our observing procedure was only $\approx 5\arcsec$.  We therefore
designed an algorithm to determine the absolute positions that were
observed based on our \ha\ image of NGC~2976, which is displayed in
Figure \ref{haimage}.  We sampled the \ha\ image with simulated
``fibers'' of the same size and location as the DensePak fibers, and
added up the flux in each simulated fiber.  By cross-correlating this
set of photometric fluxes with the observed spectroscopic fluxes
(integrated over the \ha\ line), we could measure the similarity
between them.  We repeated this process at a grid of positions around
the expected pointing center and searched for the highest value of the
cross-correlation function.  We estimate that the accuracy of the
positions derived with this method is 1\arcsec.  The algorithm failed
for one field, because only 15 of its fibers contained detectable
signal, and very little emission was visible at that location in the
image.  For this field we assumed that the offset from the expected
position was the same as the one we measured for the preceding
exposure.  Since this field is located $\sim2$\arcmin\ from the center
of the galaxy, an error of a few arcseconds in its position is
unlikely to be important.  For the other 15 fields, the algorithm gave
a smooth, well-defined peak with a cross-correlation coefficient
between 0.81 and 0.996.  We verified the results of the
cross-correlation by finding the location of the minimum rms
difference between the photometric and spectroscopic fluxes.  This
position was always within 1\arcsec\ of the cross-correlation maximum.
Fourteen of these 15 fields are located within 6\farcs9 of their
expected positions, and the other differs by 11\farcs3.

\subsection{CO Observations and Reductions}

\begin{figure}
\figurenum{1}
\epsscale{0.64}
\plotone{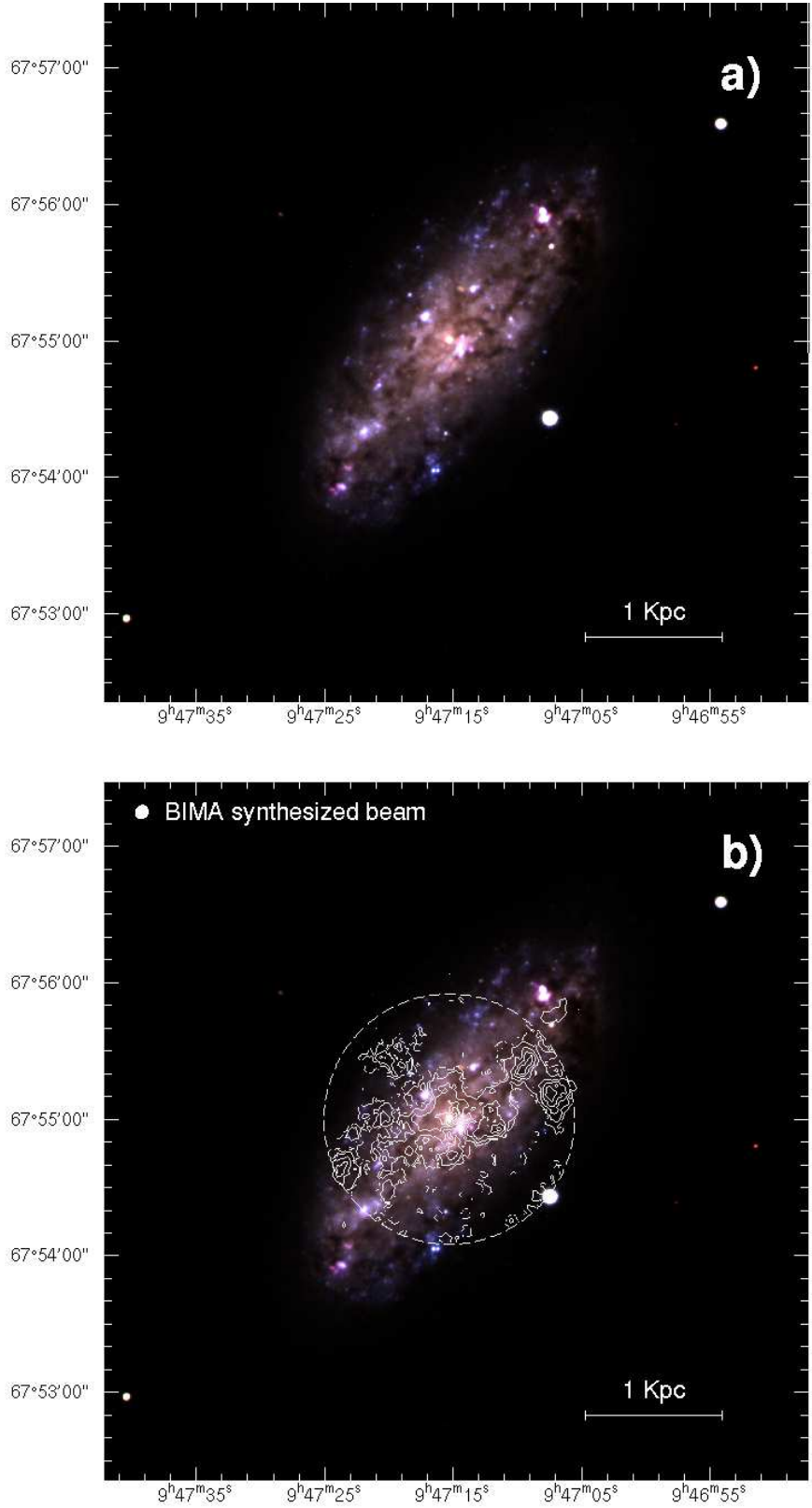}
\caption{(a) BVR composite image of NGC~2976 from the 1.8 m telescope
at Lowell Observatory.  Exposure times were 10 minutes in B and 5
minutes in V and R.  Note the distinct lack of a bulge, a bar, or any
spiral structure.  (b) BVR composite image of NGC~2976 with integrated
intensity CO contours overlaid.  Note how well the CO traces out the
optical dust lanes.  The dashed circle shows the extent of the BIMA
primary beam.  The contour levels are 0.35, 0.70, 1.4, 2.1, and 2.8
\jykms\ inside the primary beam, and a single contour at 1.4 \jykms\
is shown outside the primary beam.  For these observations, 0.35
\jykms\ corresponds to a molecular hydrogen column density of $2
\times 10^{20}$ \percmsq\ (assuming that the Galactic CO-\htwo\
conversion factor is valid in NGC~2976).  The BIMA synthesized beam
($5\farcs2 \times 6\farcs0$) is shown in the upper left corner.
\label{colorimage}}
\end{figure}

\begin{figure}
\figurenum{2}
\epsscale{0.5}
\plotone{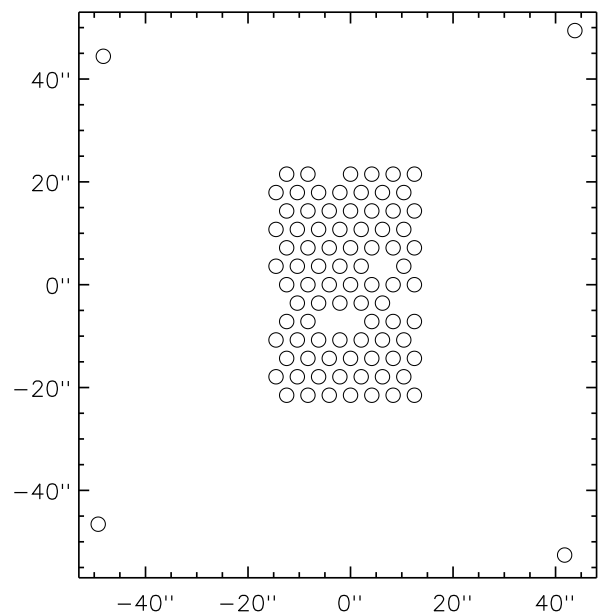}
\caption{DensePak fiber layout.  The four outlying fibers are the
sky fibers.  Since they are located only about 1\arcmin\ from the main
array, in some cases the sky spectra were contaminated by emission from
the target galaxy.
\label{densepak}}
\end{figure}

\begin{figure}
\figurenum{3}
\epsscale{1.0}
\plotone{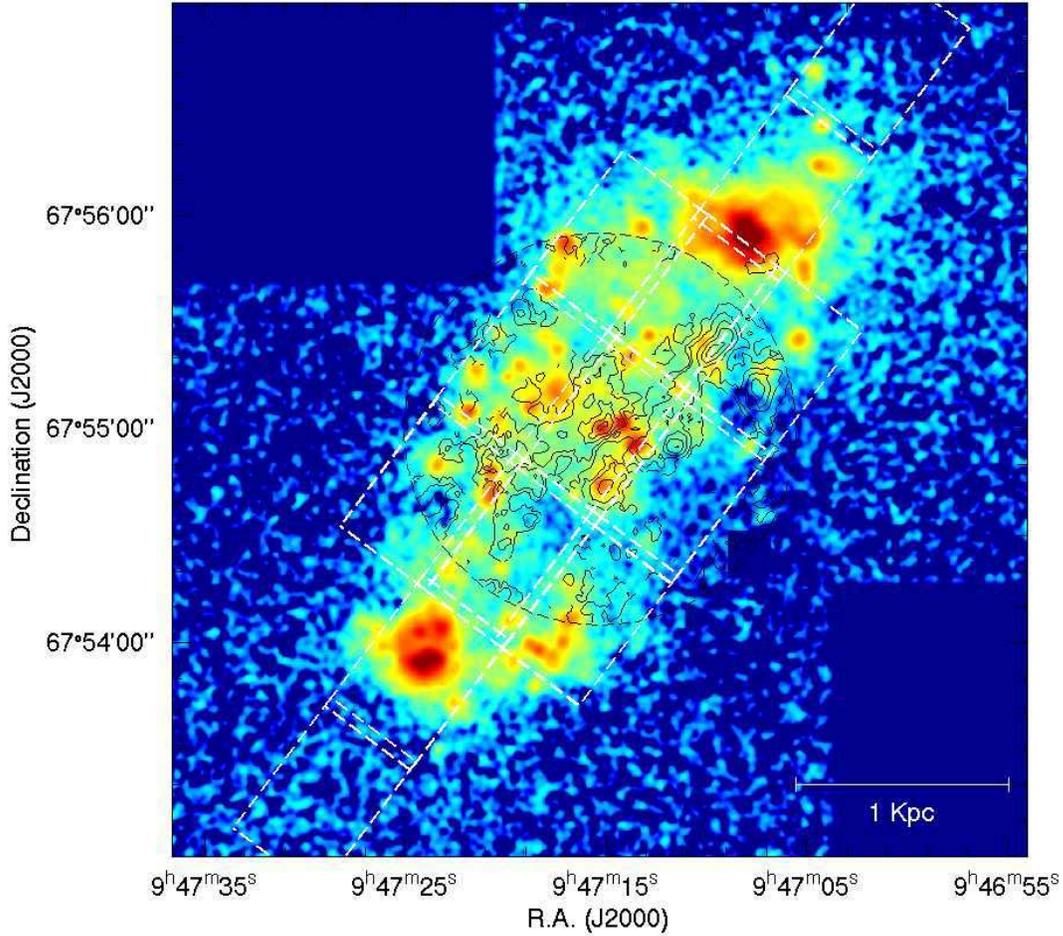}
\caption{Continuum-subtracted \ha\ image of NGC~2976.  This $4\arcmin
\times 4\arcmin$ image consists of two $1200 s$ exposures on the
Lowell 1.8 m telescope that have been combined to cover the whole
galaxy.  The images were taken through a 32 \AA-wide filter, and the
continuum was removed by appropriately scaling and subtracting images
taken through a narrow-band filter centered at 6441 \AA.  The black
contours represent integrated CO intensity, as in Figure
\ref{colorimage}b.  The white dashed rectangles overlaid on the image
show the intended locations of our DensePak pointings (these have not
been corrected for pointing errors; see Section \ref{hadata} and
Figure \ref{velfield}a), with one row or column of fibers overlapping
between every pair of adjacent pointings.  The artifacts at \ad =
$(09\hr46\min54\sec,67\degr56\arcmin35\arcsec)$ and \ad =
$(09\hr47\min07\sec,67\degr54\arcmin26\arcsec)$ are caused by masking
out residuals from bright stars.
\label{haimage}}
\end{figure}

\begin{figure}
\figurenum{4}
\epsscale{1.0}
\plotone{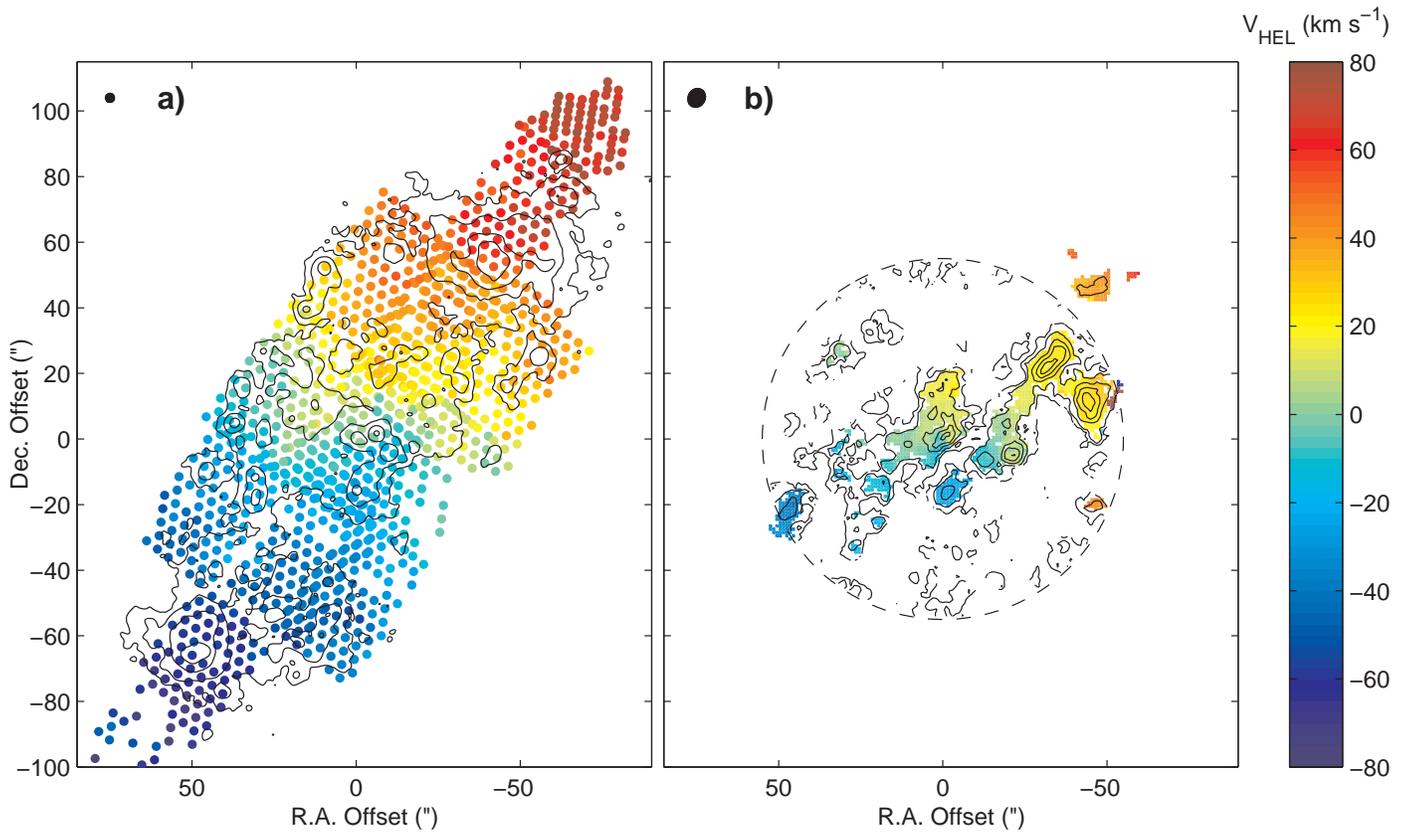}
\caption{(a) \ha\ velocity field from DensePak observations.  The
contours represent \ha\ intensity from the image displayed in Figure
\ref{haimage}.  (b) CO velocity field from BIMA observations.  The
contours represent integrated CO intensity, as in Figure
\ref{colorimage}b.  The angular resolution of each dataset is shown in
the upper left corners.
\label{velfield}}
\end{figure}

Our \co\ \jone\ observations were acquired using the B, C, and D
configurations of the 10-element BIMA array \citep{w96} between April
2001 and March 2002.  The total integration time was $\sim80$ hours,
much of which was in the most extended (B) configuration.  The BIMA
primary beam has a half-power diameter of $\sim100\arcsec$, and we
found CO emission spanning this entire width, including a cloud
outside the primary beam at $r = 70\arcsec$ (see Figure
\ref{colorimage}b).  For our observations, the spectrometer was
configured with 2 \kms\ wide channels and a 260 \kms\ bandpass.  The
individual tracks were calibrated, combined, imaged, and deconvolved
using the {\sc clean} algorithm within the MIRIAD package. The tracks
were then combined with natural weighting to create a $5\farcs2 \times
6\farcs0$ ($87\times100$ pc) synthesized beam with a position angle
(PA) of $-31\degr$.  The rms noise of the individual planes of the
datacube is 24 mJy beam$^{-1}$ in each 2 \kms\ channel.  An integrated
intensity contour map is displayed in Figure \ref{colorimage}b, and a
first moment map produced from a masked version of the datacube is
shown in Figure 5b.  Because the signal in a single channel was
relatively weak, we used the first moments to represent the velocity
at each position instead of attempting to fit Gaussians to the line
emission.  Typical uncertainties in the line velocities are 3 \kms,
and typical linewidths are 10-15 \kms\ across most of the galaxy,
although some lines are as wide as 35 \kms\ near the center.

\subsection{Optical and Near-IR Imaging and Reductions}

We observed NGC~2976 with B, V, R, and I filters at the 1.8 m Perkins
Telescope at Lowell Observatory on the photometric night of 2002
February 11.  The detector was a $2048^{2}$ Loral CCD with 15 $\mu$m
pixels and a 3\farcm2 field of view, and the seeing was
$\approx1\farcs4$.  We used exposure times of 600 s in B and 300 s in
V, R, and I and observed three overlapping positions to cover the full
extent of the galaxy.  A three-color composite of these images is
displayed in Figure \ref{colorimage}a.  To extend our set of images to
the near-infrared, we used the 2MASS JHK$_{\mbox{s}}$ Atlas images of
NGC~2976.  The 2MASS images are $8\farcm5 \times 17\arcmin$ and have
1\arcsec\ pixels, adequately sampling the $\approx3\arcsec$ seeing.

The optical data reduction, done in IDL, consisted of the following
steps: overscan subtraction, dark subtraction, flatfielding, and
cosmic ray removal.  Several bad columns were fixed by adding or
subtracting a constant so that their median values matched those of
the surrounding columns; except for the constant offset, the fluxes in
these columns do not appear to be systematically affected.  The three
images in each filter were then shifted and coadded.  We observed
several \citet{landolt92} standard fields for photometric calibration,
which was done with the IRAF implementation of {\sc daophot}
\citep{stetson87}.  With the new standard stars in these fields
identified by \citet{stetson00} in addition to the original Landolt
ones (we used Stetson's magnitudes for all of the stars), we had 34 -
38 standard star measurements per filter.  Our photometric solutions
were derived from a least-squares fit to the following formula:

\begin{equation}
m = m_{\mbox{\small{instr}}} + C + f (V - I) + g (a - 1) \ ,
\end{equation}

\noindent where $m$ is the apparent magnitude,
$m_{\mbox{\small{instr}}}$ is the instrumental magnitude ($25 - 2.5
\log{\mbox{flux}} + 2.5 \log{\mbox{integration time}}$), $C$ is a
constant that sets the instrumental zero point, $f$ is the color
coefficient, $V - I$ is the color of the object, and $g$ is the
extinction coefficient.  Our observations did not span a large enough
range of airmass to determine the extinction coefficient directly,
so we used previously derived values.  Reasonable ranges for the
coefficients are $0.2<g_{B}<0.4$ mag airmass$^{-1}$, $0.1<g_{V}<0.3$
mag airmass$^{-1}$, $0.05<g_{R}<0.15$ mag airmass$^{-1}$, and
$0.02<g_{I}<0.12$ mag airmass$^{-1}$ (P. Massey, private
communication).  For B and V, we used values of 0.27 mag
airmass$^{-1}$ and 0.15 mag airmass$^{-1}$, which were the mean values
of $\sim15$ measurements made between 1997 and 1999 at the same
telescope (D. Hunter, private communication).  Lacking comparable
measurements in R and I, we used the standard Lick Observatory values
of 0.11 mag airmass$^{-1}$ and 0.08 mag airmass$^{-1}$, respectively.
These fall within the reasonable ranges for both filters, and the
Hunter B and V measurements are very close to the Lick values.  Since
all of our images were taken at airmasses close to 1.2, these
assumptions are unlikely to cause significant errors.

In order to double-check our photometric solutions, we obtained the V-
and I-band Keck\footnote{The W.~M. Keck Observatory is operated as a
scientific partnership among the California Institute of Technology,
the University of California, and the National Aeronautics and Space
Administration. The Observatory was made possible by the generous
financial support of the W.M. Keck Foundation.} images that
\citet{mendez} acquired for the purpose of measuring the TRGB distance
to NGC~2976.  These images were taken with the Low-Resolution Imaging
Spectrometer \citep{oke95}, and cover a $5\arcmin \times 7\arcmin$
field.  Exposure times were 300 s in I and 400 s in V.

\subsection{Surface Brightness Profiles}
\label{sbprofsec}

We used the IRAF routine {\sc ellipse} in the STSDAS package to
perform surface photometry on the images.  The routine fits elliptical
isophotes to a galaxy image at specified radii, and allows the
position angle (PA), ellipticity, and center to change with radius.
There is no evidence that the PA changes with radius, so we used the
average value of 143\degr, identical to the cataloged PA of the galaxy
\citep[hereafter RC3]{rc3}.  The ellipticity $\epsilon$, which is
related to the inclination angle via the formula $\cos{}^{2}i =
[(1-\epsilon)^{2} -
(1-\epsilon_{max})^{2}]/[1-(1-\epsilon_{max})^{2}]$, where $\epsilon
\equiv 1-b/a$, $a$ and $b$ are the major and minor axis lengths, and
$\epsilon_{max} = 0.8$, varies from $\sim0.4$ to $\sim0.7$ in the
inner part of the galaxy before converging to a constant value of 0.49
for $r > 114\arcsec$.  Galaxies often display such behavior, and it is
not generally interpreted as a changing inclination angle with radius.
Accordingly, we use an ellipticity of 0.49 for the whole galaxy.  The
corresponding inclination angle is 61.4\degr, the same as the RC3
inclination of 61.5\degr\ within the uncertainties.  The center of the
isophotal fits changed incoherently with radius before converging for
$r > 120\arcsec$.  The isophotal center was within a few arcseconds of
the visually obvious nucleus at \ad = (09\hr47\min15.3\sec,
67\degr55\min00.4\sec), so we used the nucleus as the fit center.  The
coordinates of the nucleus coincide with the cataloged galaxy
positions within their uncertainties \citep{ugc,uzc}.  We then ran
{\sc ellipse} again with all of the parameters fixed to produce the
final surface brightness profiles.  We fit ellipses every 2\arcsec\ 
out to a radius of 172\arcsec, where the ellipses began to run off the
edge of the image.

We also ran {\sc ellipse} on the Keck images with the same parameters.
This revealed some systematic differences between the Lowell and Keck
photometry: although the profile shapes were quite similar in the two
datasets, the Lowell V magnitudes are 0.1 mag brighter than the Keck V
magnitudes, and the Lowell I magnitudes are 0.1 mag fainter than the
Keck I magnitudes.  The cause of this discrepancy is not clear, and it
is worrisome because a 0.2 mag change in the galaxy color is
significant.  However, as we will show in \S \ref{stars}, the measured
Lowell colors all predict stellar mass-to-light ratios that are
consistent with one another, while the Keck $V-I$ color predicts a
noticeably higher mass-to-light ratio that is inconsistent with the
other determinations.  An additional piece of evidence in favor of the
Lowell magnitudes is that the tabulated $B-V$ color in, e.g., the RC3
is close to our measured value, so we conclude that it is safe to
assume that our Lowell photometry is accurate.  Since the LRIS field
of view is larger than that of the Lowell CCD, we also used the Keck
images to verify that the light profile does not change shape at
larger radii, and to measure the fraction of the total flux that we
missed due to the limited extent of the Lowell mosaic.  We estimate
that $\sim96 \%$ of the galaxy's light is contained within the $r =
172\arcsec$ ellipse out to which we measured, so our integrated
magnitude measurements should probably be revised upwards by 4~\%
(0.04 mag).

The measured surface brightness profiles are corrected by applying the
\citet{sfd98} Galactic extinction estimates in each band.  To account
for extinction within NGC~2976, we used an inclination-based approach,
as described by \citet{sakai00}.  \citet{sakai00} give internal
extinction coefficients for all of the bands we use except J, so to
determine the J-band correction we interpolated their results from the
other bands and found that $A_{J} = 0.8 A_{I}$.  For our best-fit
axial ratio of 1.96, we estimate that the internal extinction in
magnitudes in the seven bands is (from B to K$_{\mbox{s}}$): 0.23,
0.20, 0.18, 0.13, 0.11, 0.07, and 0.03.

\begin{figure}
\figurenum{5}
\epsscale{1.0}
\plotone{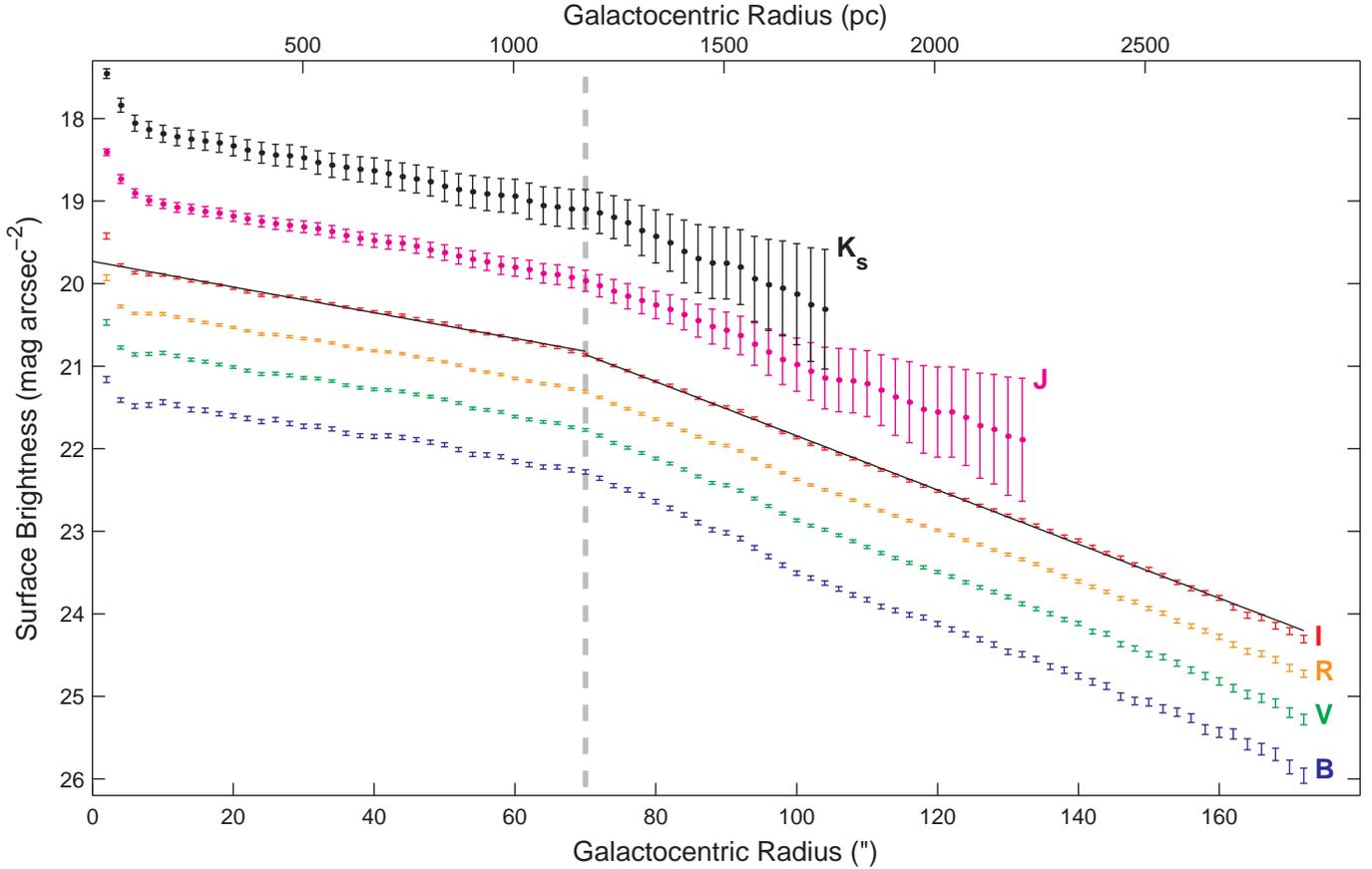}
\caption{Optical and near-infrared surface brightness profiles of
NGC~2976.  For the J and K$_{\mbox{s}}$ profiles we plot data points
and error bars, but we omit the points for B, V, R, and I because they
would obscure the error bars.  The J and K$_{\mbox{s}}$ data can be
traced further out, but we do not plot the data beyond where the
uncertainties reach a factor of 2 (0.75 mag).  The H-band profile has
also been left off for clarity; the error bars for H and
K$_{\mbox{s}}$ overlap at most radii.  In each color, the nucleus,
exponential inner disk, and exponential outer disk are all visible.
In the optical filters, there is a transition region between the inner
and outer disks where the colors are bluer than the disk values.
Exponential fits to the I-band profile are shown by the solid black
lines.  The vertical dashed line at a radius of 70\arcsec\ emphasizes
the breakpoint between the inner and outer disks.  For central surface
brightnesses and disk scale lengths, see Table 1.
\label{sbprofs}}
\end{figure}

The surface brightness profiles, displayed in Figure \ref{sbprofs},
are qualitatively similar in all of the filters.  NGC~2976 clearly
contains three components: a nucleus, an exponential inner disk, and
an exponential outer disk.  Since the nucleus is not resolved in any
of our images, we used the HST/NICMOS images acquired by
\citet{boker99} to estimate that its radius is less than 0\farcs36 (6
pc).  It seems to be too reddened to reliably derive a mass-to-light
ratio from its colors.  The nuclear luminosity is $6 \times 10^{6}
L_{K,\odot}$, so the rotation velocity due to the nucleus is $39
(M_{nuc} / L_{K})^{1/2} (r / 1\arcsec)^{-1/2}$ \kms, where $M_{nuc} /
L_{K}$ is the stellar mass-to-light ratio of the nucleus in solar
units.  If we assume that the mass-to-light ratio is the same as the
maximum allowable value for the disk (see \S \ref{dmlimits}), the
nucleus becomes dynamically insignificant outside 10\arcsec.  Because
the nucleus is probably a large cluster of young stars (judging by its
compactness, luminosity, and \ha\ emission), its actual mass-to-light
ratio is likely much lower.  Parameters for the disk of NGC~2976 in
each band are listed in Table \ref{disktab}.  The presence of an outer
exponential disk, with a surface brightness that declines more quickly
than would be expected from extrapolating the inner disk, has been
seen in other spiral galaxies \citep[][hereafter
PDLA]{nj97,pohlen01,pohlen02}.  The ratio of the inner scale length to
the outer scale length is 2.1, consistent with the value of $2.0 \pm
0.2$ measured by PDLA for four other galaxies.  In fact, NGC~2976 only
differs from the galaxies PDLA observed in that the break between the
inner and outer disks occurs close-in, at 1 inner disk scale length
instead of $\sim4$ scale lengths.  NGC~2976 is an order of magnitude
less luminous than the galaxies in the PDLA sample, suggesting that
the break radius might be a function of luminosity.

\section{BARYONIC COMPONENTS OF NGC~2976}
\label{baryons}

Because our images of NGC~2976 do not reveal a bulge or a bar, and its
nucleus is dynamically unimportant, the only relevant reservoirs of
baryons to consider are the stellar and gaseous disks.

\subsection{The Stellar Disk}
\label{stars}

There are two obvious approaches to studying the importance of the
stellar contribution to a galaxy rotation curve: 1) compare multicolor
surface photometry of the galaxy with the predictions of stellar
population synthesis models to obtain an estimate of the stellar
mass-to-light ratio (\msl) that is independent of the galaxy
kinematics, or 2) leave \msl\ as a free parameter while simultaneously
fitting a scaled stellar disk and a dark matter halo to the observed
rotation curve.  The second technique has a very significant drawback:
$\chi^{2}$ is insensitive to changes in \msl\ during rotation curve
fits \citep{mgdb98,swaters99,us}, so the fit with the lowest value of
$\chi^{2}$ does not necessarily convey any information about the value
of \msl.  As an illustration of this effect, the best fit often turns
out to be $\msl = 0$, even though that is clearly not correct.  We
would therefore like to have an independent constraint on the
mass-to-light ratio so that we do not have to leave it as a free
parameter.  For this reason, and because we have BVRIJHK$_{\mbox{s}}$
photometry available, we choose the first method.

\subsubsection{Population Synthesis Constraints On \msl}
\label{popsynth}

One way to estimate a stellar mass-to-light ratio from photometry
alone is to use the semi-empirical relationships derived by
\citet{bdj01}.  \citet{bdj01} showed that the colors of spiral
galaxies are strongly correlated with the mass-to-light ratios of
their stellar populations.  With our multicolor photometry, we can
construct the entire array of colors for which they give formulas, and
then calculate the expected mass-to-light ratios, which are listed in
Table \ref{mltable}.  The average predicted values from the six
tabulated inner disk colors are $0.48 \pm 0.02 \mlk$ in K
band\footnote{We will mostly use the K-band stellar disk for the
remainder of the paper for the following reasons: 1) K-band light is
the best tracer of the stellar \emph{mass} distribution and the least
skewed by luminous young stars from recent star formation, and 2) K
band is the least affected by extinction.}, and $1.07 \pm 0.07 \mlr$
in R band.  That all of the colors predict consistent mass-to-light
ratios is an indication that the predictions have some validity for
this galaxy.  It must be noted, however, that the \citet{bdj01}
mass-to-light ratios are derived assuming that galaxies have maximal
disks.  If the average galaxy has a disk that is a factor $f$ ($0 \le
f \le 1$) less than maximal, then the predicted mass-to-light ratios
from their calculations must also be scaled by the same factor $f$.  A
further uncertainty in this analysis is the initial mass function
(IMF), which may not follow the assumed scaled Salpeter form,
particularly at low masses.

An alternative method to measure \mslk\ photometrically is to compare
the observed colors directly to the outputs of publicly available
stellar population synthesis models.  We used the Starburst99
population synthesis models \citep{l99} to attempt to constrain \mslk\
in this way.  For a given star formation history (constant star
formation rate or instantaneous burst of star formation), Starburst99
predicts colors and luminosities as a function of time.  This allows
us to search systematically for the population age that matches the
observed set of colors most closely.  The two best Starburst99 models
are 1) a population with a small (10 \%) young component that has been
forming stars continously, and the remaining stars in an old ($t
\gtrsim 3$ Gyr) population that formed in an instantaneous burst, and
2) a very young ($\sim 10^{7}$ yr old) starburst.  However, the
mass-to-light ratios of these models seem rather implausible: $\mslk >
2$ for model 1) and $\mslk \approx 0.02$ for model 2).  Neither of
these values is compatible with observed values of $\mslk$ in the few
other galaxies for which measurements are available
\citep[e.g.,][]{om01,vbb02} or with the predictions of \citet{bdj01}.
We also tried the online population synthesis code described by
\citet{worthey94}, which predicts colors and mass-to-light ratios for
an arbitrary combination of input stellar ages and metallicities.  It
is more difficult to do a comprehensive search through the likely
parameter space with this technique, but we did find that a mixture of
70~\% of an old (but metal-rich) population and 30~\% of a young
population ($t = 1.1$ Gyr) comes close to reproducing the observed
colors (assuming a Miller-Scalo IMF), yielding a mass-to-light ratio
of 0.31 \mlk.

We conclude that it is not possible to uniquely determine the \mslk\
for NGC~2976 by comparing the galaxy colors with the predictions of
current population synthesis models.  From this information alone,
NGC~2976 could contain either a very young starburst, or a normal,
mixed stellar population with a low to moderate (0.3 to 0.5) \mslk.
There are three reasons for discounting the starburst possibility in
NGC~2976.  First, the colors of the outer disk of NGC~2976, where our
observations show no evidence for widespread star formation, are very
similar to the inner disk colors, particularly in the near-infrared
where extinction is less important (as shown in Table \ref{mltable}).
It is unlikely that the galaxy contains a starburst and an old
population that coincidentally have the same observed colors.  Second,
the starburst would have to be unusually young (in which case our
observations of it at this particular time are rather fortuitous), and
also quite strong, dominating not only the light output from the
galaxy, but also containing a significant fraction of its total
stellar mass (otherwise the mass-to-light ratio would begin to run
into the kinematic limit; see \S \ref{dmlimits}).  And finally, the
visual appearance of the galaxy is not suggestive of a vigorous
starburst.  The more likely alternative is that NGC~2976 has a
substantial old component to its stellar population, driving \mslk\
toward the values of a few tenths that are seen in other galaxies.

\subsubsection{Rotation Velocities Due to a Thin Disk}
\label{thindisk}

In order to compare the stellar rotation velocities to the observed
rotation curve, we calculate the rotation velocities for material
confined to a thin disk.  Because the disk of NGC~2976 is not a pure
exponential, its rotation curve must be calculated numerically.  We
perform the calculation with the routine {\sc ccdpot}, which is based
on a derivation given by \citet{bt87}, in the NEMO package
\citep{teuben}.  This rotation curve is similar to that from the
fitted exponential disk out to the breakpoint between the inner and
outer disks, where it briefly exceeds the exponential disk rotation
curve, and then begins to decline more quickly (as expected, since the
surface density at large radii is lower than in the single exponential
case).  Our calculations assume an infinitely thin disk for
simplicity; allowing the disk to have some thickness leaves the shape
of the rotation curve almost unchanged, but lowers its amplitude, thus
raising the allowed \mslk\ \citep{swaters99,ppcl02}.  For a scale
height equal to 1/6 of the disk scale length, the rotation curve is
lowered by about 10~\% \citep{ppcl02}, so that the allowed \mslk\ may
be 20~\% higher than in the infinitely thin case (since $v_{rot}
\propto \sqrt{\mslk}$).

\subsection{The Gas Disk}
\label{gas}

\begin{figure}
\figurenum{6}
\epsscale{0.5}
\plotone{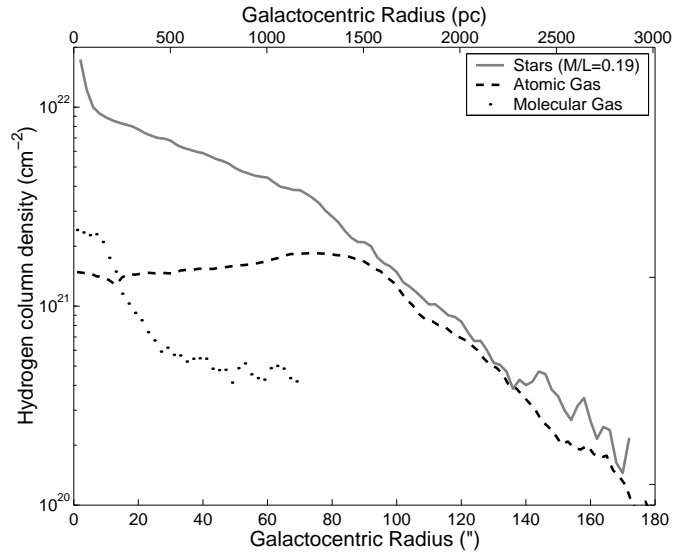}
\caption{Surface densities of the stars and gas in NGC~2976.  The \hi\
and \htwo\ surface densities do not include helium, so the stellar
surface densities are divided by a factor of 1.3 to match.  Of the
baryonic components, the stars dominate the inner disk, but the \hi\
is almost as important in the outer disk.  The molecular gas surface
density outside 40\arcsec\ is quite uncertain.
\label{surfdens}}
\end{figure}

The atomic and molecular gas disks of NGC~2976 do not contribute
appreciably to its measured rotation curve.  Although it is rich in CO
for a dwarf galaxy, the measured total flux of $\sim45$ \jykms\ over
the central 750 pc \citep{fcrao} implies only $\sim10^{7} M_{\odot}$
of molecular gas (including helium), if the Galactic CO-\htwo\
conversion factor is valid in NGC~2976.  The total molecular mass
might be somewhat larger, because the BIMA observations did not cover
the large \hii\ regions at either end of the inner disk, which are
likely associated with molecular clouds.  Nevertheless, the molecular
material is not dynamically significant globally or locally,
regardless of how it is distributed.  The atomic gas mass is much
larger, at $1.5 \times 10^{8} M_{\odot}$ \citep*{ads81,si02a}.  We
adapt the \hi\ surface density distribution from the data presented by
\citet{si02a}.  The stellar, atomic, and molecular surface densities
are plotted in Figure \ref{surfdens}.  Even with a low \mslk, the
stars are clearly the dominant reservoir of baryons in NGC~2976.  It
is noteworthy that the \hi\ and stellar scale lengths in the outer
galaxy appear nearly identical, and the surface densities are
comparable as well.  We calculate the rotation curves of the gaseous
components directly from their surface density profiles (again
assuming zero thickness) using the same method as for the stars.

\section{ROTATION CURVE AND DARK MATTER HALO OF NGC~2976}
\label{darkmatter}

Now that we have a handle on the behavior of the stellar and gas disks
of the galaxy, we can move on to our primary goal of constraining the
structure of the dark matter halo.  First, we convert our
two-dimensional velocity field into a one-dimensional rotation curve.
This is accomplished by fitting tilted-ring models to the velocity
field using three complementary techniques.  The algorithms are
mentioned briefly below, and more detailed descriptions are given in
Appendix \ref{algorithms}.  {\sc Rotcur} breaks the velocity field
into rings and fits for the PA, inclination, center, systemic
velocity, and rotation velocity in each ring.  {\sc Ringfit} also
divides the galaxy into rings, and it fits for the rotation velocity,
the radial velocity (in the plane of the galaxy), and the systemic
velocity in each ring.  {\sc Ringfit} thus has the desirable feature
that a simple form of noncircular motions are included in the fit.
The third algorithm, \rotcurshape, fits the entire velocity field with
a single PA, inclination, center, and systemic velocity, and also
assumes a functional form for the rotation curve and solves for the
parameters of that function.  Fit results are similar for all three
procedures, although fitting for radial velocities in addition to
rotation does make the rotation curve somewhat shallower (see \S
\ref{rcsystematics}).

\subsection{Rotation Curve of NGC~2976}
\label{rcsec}

It is apparent from the data (Figure \ref{velfield}) that the
velocity field near the center of the galaxy cannot be adequately
described by rotation alone.  There are two choices for how to
proceed: 1) use additional Fourier terms to describe the velocity
field, or 2) allow for changes in the position angle of the galaxy
with radius.  The second possibility, which is difficult to reconcile
with the photometry, is discussed in \S \ref{noncirc}; for now, we
will use Fourier analysis to provide an accurate description of the
velocity field.  The next Fourier term beyond pure rotation
($\cos{\theta}$, where $\theta$ is the angle from the major axis in
the plane of the galaxy; see Appendix \ref{algorithms}) is pure radial
motions ($\sin{\theta}$).  We have investigated the decomposition of
the velocity field using higher order terms, and found that they are
much smaller than the rotation and radial components and are
consistent with noise. Therefore, we leave those terms out of our
subsequent analysis. 

\begin{figure}
\figurenum{7}
\epsscale{1.0}
\plotone{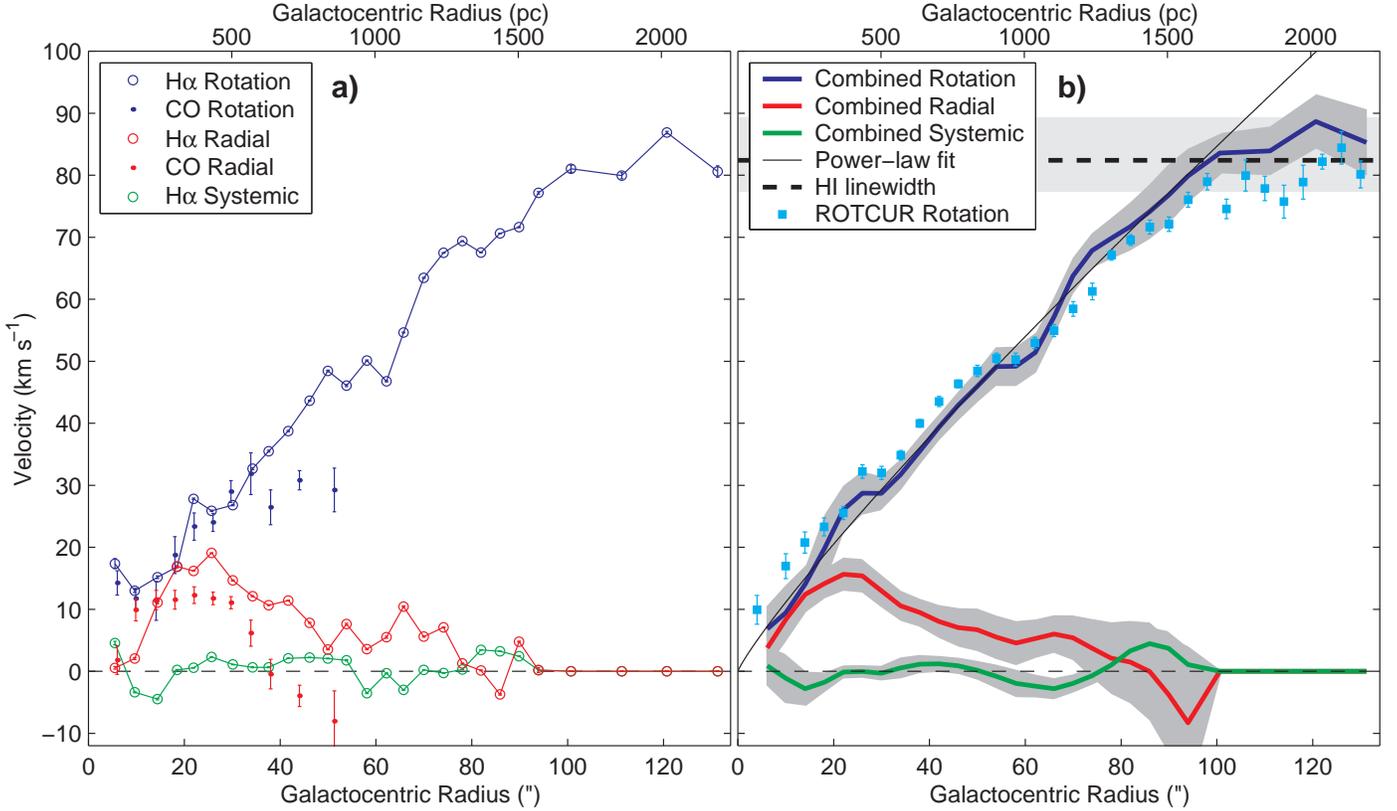}
\caption{(a) \ha\ and CO velocity field decompositions from {\sc
ringfit}.  The blue points represent rotation velocities, the red
points represent radial velocities, and the green points are systemic
velocities.  Open symbols are from the \ha\ velocity field, and filled
symbols are CO data.  The rotational and radial velocities of the CO
and \ha\ are consistent with each other.  Because the number of
independent CO data points is small, we reduced the number of degrees
of freedom in the fit by fixing the systemic velocities.  The error
bars are only statistical errors, which substantially underestimate
the true uncertainties.  (b) Combined velocity field decompositions
from \ringfit.  To create this rotation curve, we combined the \ha\
and CO data into a single velocity field.  We then ran a Monte Carlo
simulation in which the velocity field is fit many times, assuming a
PA, inclination, and center position that are drawn randomly from the
Gaussian distributions $PA = -37\degr \pm 5\degr$, $i = 61.5\degr \pm
3\degr$, and center = nucleus $\pm$ 2\arcsec.  The curves show the
mean results from 1000 realizations of the simulation, and the shading
that follows the curves represents $1\sigma$ systematic uncertainties
in each of the plotted quantities.  The thin black line is a power-law
fit to the rotation curve, corresponding to a density profile of
$\rhotot \propto r^{-0.27}$.  The cyan points are the \rotcur\
rotation curve, showing the difference that arises if the radial
velocities are not included in the fit.  Note that although we have
plotted the radial motions as positive velocities, whether they
represent inflow or outflow cannot be determined without knowing which
side of the galaxy is the near side.
\label{rcfig}}
\end{figure}

\begin{figure}
\figurenum{8}
\epsscale{1.0}
\plotone{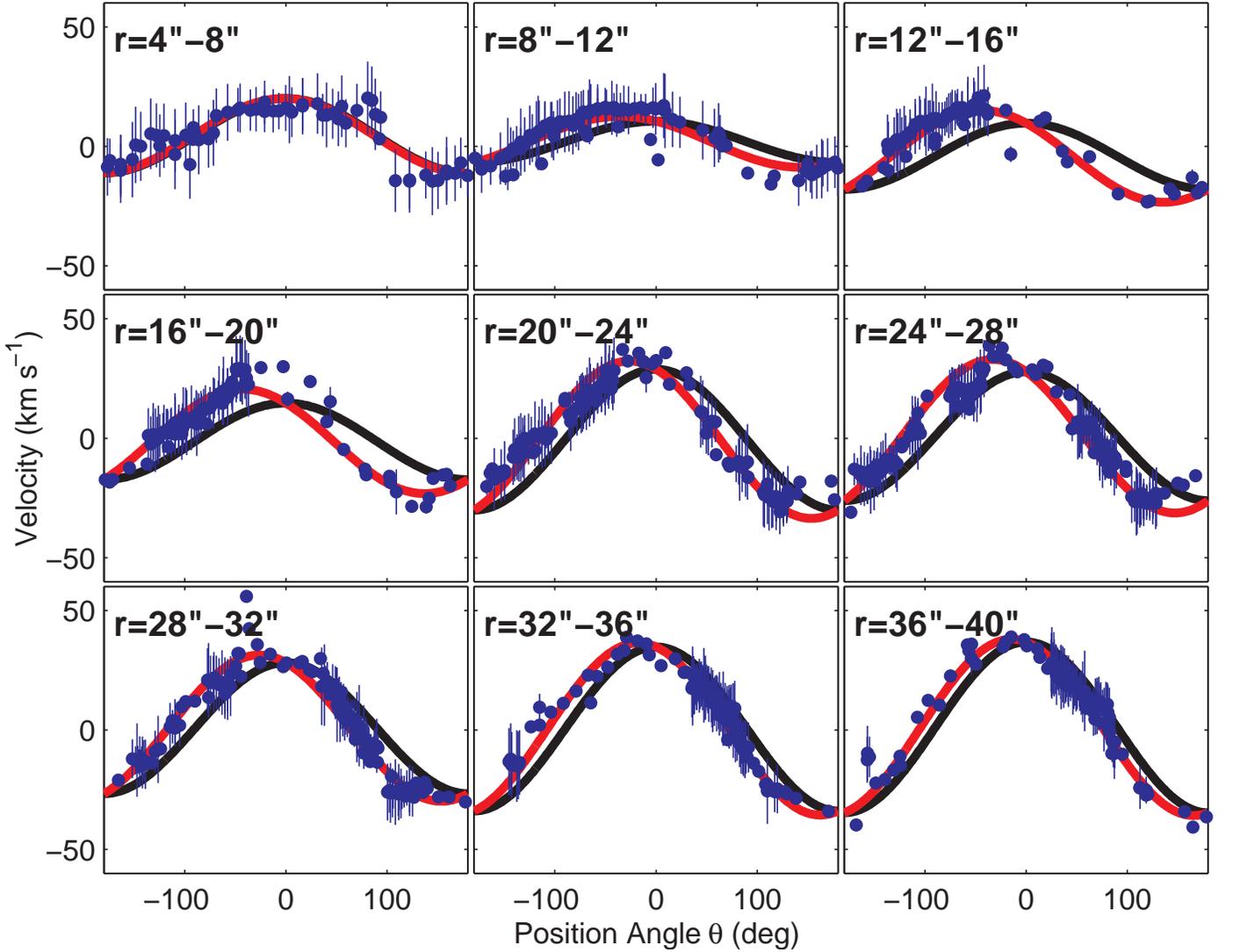}
\caption{Fits to the velocity field using \ringfit.  The observed
velocities are plotted as a function of angle $\theta$ in the plane of
the galaxy, where $\theta = 0$ is the major axis.  Data points with
small error bars are from the \ha\ velocity field (and are all
independent), and data points with large error bars are from the CO
velocity field (and are not all independent; the error bars have been
increased to account for this).  The black curves show the rotational
component of the fits ($\cos{\theta}$), and the red curves show the
fits including both rotation and radial motions ($\sin{\theta}$).  The
displacement of the velocity maxima from $\theta = 0$ illustrates the
need for radial motions in the fits.  At radii beyond 40\arcsec,
radial motions are not needed to obtain good fits to the data.  
\label{ringfitfig}}
\end{figure}

Our final rotation curve was derived with \ringfit, so that we could
account for the radial motions that are present in the velocity field.
We first fit the \ha\ and CO velocity fields separately to verify that
they agree with each other, as displayed in Figure \ref{rcfig}a.  At
small radii ($r < 40\arcsec$), it is evident that the derived rotation
curves and radial motions do agree, although they begin to diverge
somewhat at $r > 40\arcsec$.  However, the CO ring fits at these radii
are based on only one or two independent measurements, so the apparent
difference between the CO and \ha\ velocity fields is not significant.
Therefore, we combine the two datasets and fit again, weighting each
data point by the inverse square of its statistical uncertainty.  The
fits for rings at $r < 40\arcsec$ are displayed in Figure
\ref{ringfitfig}, where it is apparent that the velocity maxima do not
lie along the major axis.  This indicates the presence of radial
motions, which could not have been measured with longslit observations
or \rotcur\ fitting.  The rotation curve from these fits is plotted in
Figure \ref{rcfig}b, and the significance of the radial velocities is
again apparent.  The estimated systematic uncertainties (the
derivation of which is described in \S \ref{rcsystematics}) are shown
by the shaded gray areas surrounding each curve.

\begin{figure}
\figurenum{9}
\epsscale{0.6}
\plotone{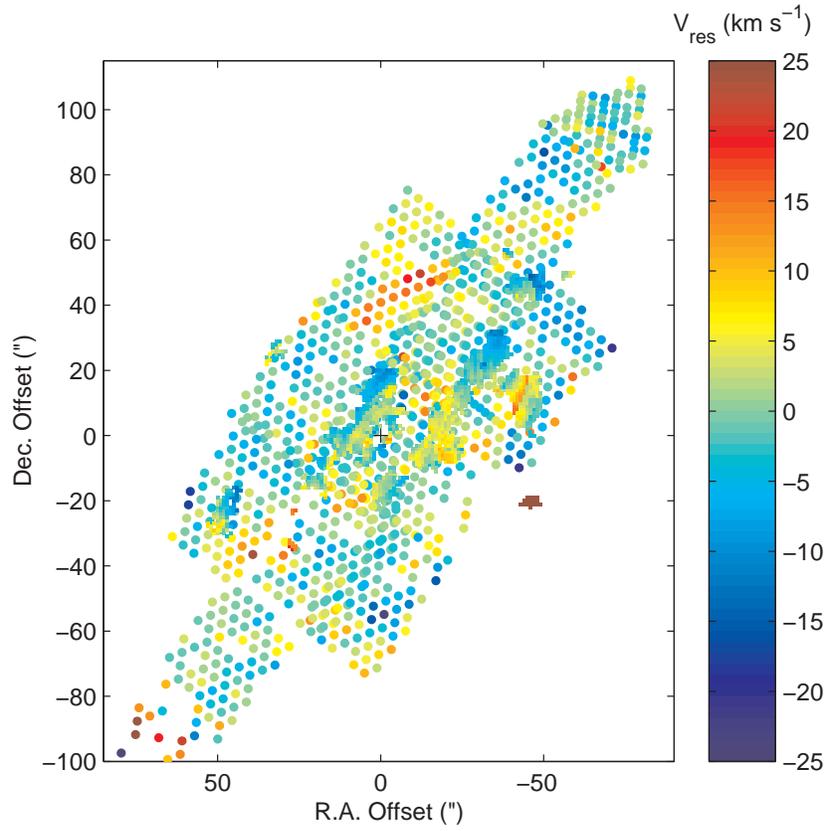}
\caption{Residual velocity field after subtracting \ringfit\ model
from the combined \ha\ and CO velocity fields.  \ha\ data are shown by
the circles, and the CO data are shown by the closely-packed square
pixels.  The rms of the residuals is 6.4 \kms; 5.5 \kms\ if the small
patch of probably spurious CO emission southwest of the galaxy at
(R.A. offset, Dec. Offset) = (-45\arcsec, -20\arcsec) is excluded.
\label{resids}}
\end{figure}

The residual velocity field after subtracting this best-fit model is
displayed in Figure \ref{resids}.  Although individual residuals are
occasionally as large as 30 \kms, the rms of the residual field is 6.4
\kms, and there are no obvious systematic trends.  The random velocity
variations in the residual map are much larger than the uncertainties
in the observed velocities, and the value of 6.4 \kms\ is consistent
with the random velocities of gas observed in other galaxies; the
residuals therefore likely represent real small-scale structure in the
velocity field.  The rotation velocities and radial velocities with
their associated uncertainties, and the stellar and gas rotation
curves, are all listed in Table \ref{data}.  In order to incorporate
more accurately the uncertainties in the rotation curve, the values
listed in Table \ref{data} and plotted in Figure \ref{rcfig}b are the
mean values obtained from a Monte Carlo study rather than directly
from the data (see \S\ref{dvrot}).

\begin{figure}
\figurenum{10}
\epsscale{1.0}
\plotone{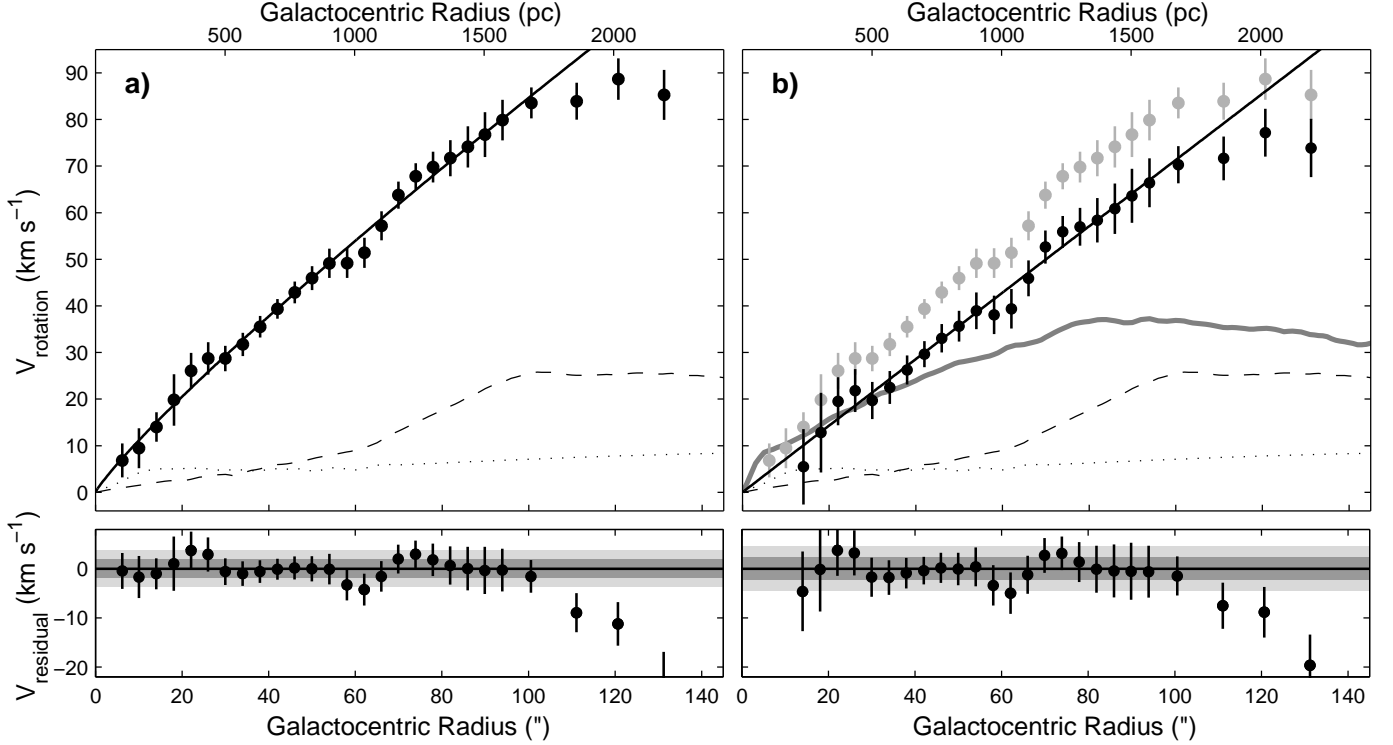}
\caption{(a) Minimum disk rotation curve of NGC~2976.  Here we assume
that the dark matter is dynamically dominant over the baryons at all
radii, so that the observed rotation velocities (black circles) are
attributable entirely to the dark matter halo.  This represents the
cuspiest possible shape for the dark matter halo.  The plotted
error bars are combined statistical and systematic uncertainties.  The
rotation velocities due to \hi\ and \htwo\ are plotted as dashed and
dotted curves, respectively.  A power law fit to the rotation curve is
shown by the solid black curve.  The corresponding density profile is
$\rho \propto r^{-0.27}$.  Residuals from the fit are displayed in the
lower panel, and $1\sigma$ and $2\sigma$ departures from the fit are
represented by the shaded regions.  (b) Maximum disk rotation curve of
NGC~2976.  In this case, we scale up the stellar disk (solid gray
curve) as high as the observed rotation velocities (gray circles)
allow.  The stellar disk shown here has a mass-to-light ratio of 0.19
\mlk.  This is the most massive stellar disk that can be present
without making the dark matter density increase with radius, which is
probably not physically realistic.  After subtracting the rotation
velocities due to the stars, the rotation velocities due to the \hi\
(dashed curve), and the rotation velocities due to the \htwo\ (dotted
curve) in quadrature from the observed rotation curve, the dark matter
rotation velocities are displayed as black circles.  The two missing
data points near the center of the galaxy had $v_{rot} < v_{*}$,
yielding imaginary $v_{halo}$.  The solid black curve is a power law
fit to the halo velocities (for $14\arcsec < r < 105\arcsec$) which
corresponds to a density profile of $\rhodm \propto r^{-0.01}$.  The
halo residuals after the power law fit are displayed in the bottom
panel.
\label{diskfig}}
\end{figure}

The rotation curve of NGC~2976 is well-described by a power law from
the center of the galaxy out to a radius of almost 2 kpc, as displayed
in Figure \ref{diskfig}a.  The residuals after subtracting the fit
from the rotation curve are shown in the bottom panel.  The rotation
curve only begins to deviate systematically from power-law behavior at
$r \approx 110\arcsec$ (1.84 kpc).  The \emph{total} (baryonic plus
dark matter) density profile corresponding to the rotation curve is
$\rhotot = 1.6 (r/1 \mbox{ pc})^{-0.27 \pm 0.09}$~\mpc\ (see Appendix
\ref{densityprofs}, Equations \ref{alphaeq} and \ref{betaeq}, for the
conversion between power laws in velocity and density).  This density
profile is the mean of the fits to 1000 Monte Carlo rotation curves,
which represents a more accurate estimate of the uncertainties than
the fit to the single rotation curve shown in Figure \ref{rcfig}b.  In
the following subsection, we show that the density profile of the
\emph{dark matter halo alone} follows a shallower power law.

A key assumption underlying the derivation of this density profile is
that the orbits are circular, and therefore that the gravitational and
centripetal forces are in equilibrium.  This assumption is not likely
to be correct in detail, but an inversion of the velocity field
(including noncircular motions) to obtain the underlying
nonaxisymmetric potential is beyond the scope of this paper.
Nevertheless, we note that the radial motions are comparable in
magnitude to the rotation only for the inner four points of the
rotation curve (the central 300 pc of the galaxy).  At larger radii
the rotation clearly dominates, and the orbits are nearly circular.
If we fit the rotation curve using only points between 300 pc and 1.8
kpc --- where the radial motions are probably unimportant --- the
derived density profile is almost identical to the one described in
the previous paragraph.  This suggests that a more complete analysis
of the effect of noncircular motions on the inversion from a rotation
curve to a density profile should not have a large impact on the
derived slope of the density profile.

\subsection{Limits on the Dark Matter Halo}
\label{dmlimits}

To reveal the shape of the density profile of the dark matter halo, we
first remove the rotational velocities contributed by the baryonic
components of the galaxy.  The rotation curve of the dark matter halo
is defined by

\begin{equation}
v_{halo}^{2} = v_{rot}^{2} - v_{*,rot}^{2} - \vhirot^{2} -
\vmolrot^{2}.
\label{vhaloeq}
\end{equation}

\noindent
We determine the lower limit to the dark matter density profile slope
by maximizing the rotation curve contribution from the stellar disk.
The maximum possible stellar rotation curve is set by scaling up the
mass-to-light ratio of the stellar disk until the criterion

\begin{equation}
v_{*,rot}^{2} < v_{rot}^{2} - \vhirot^{2} - \vmolrot^{2}
\label{maxdiskeq}
\end{equation}

\noindent
is no longer met at every point of the rotation curve.  This
requirement sets maximum disk mass-to-light ratios of $\mslk =
0.09^{+0.15}_{-0.08} \mlk$ and $\mslr = 0.53^{+0.56}_{-0.46} \mlr$,
where the uncertainties are calculated by replacing $v_{rot}$ with
$v_{rot} \pm \delta v_{rot}$ in Equation \ref{maxdiskeq}.

\begin{figure}
\figurenum{11}
\epsscale{0.5}
\plotone{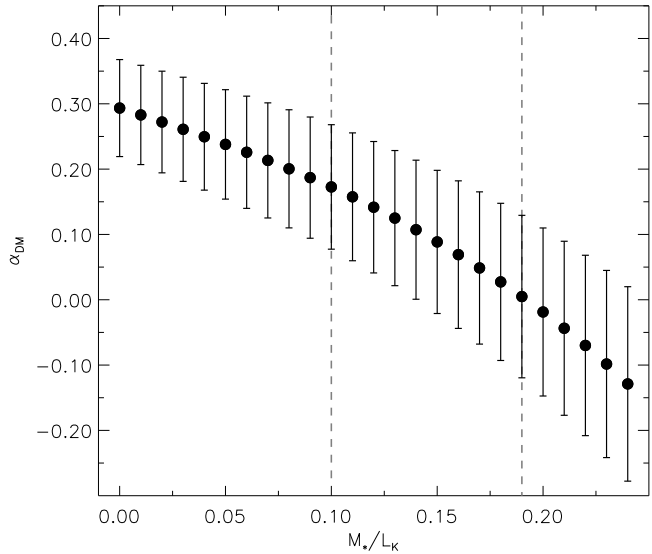}
\caption{Dark matter density profile slope $\alphadm$ as a function of
the assumed K-band stellar mass-to-light ratio.  The error bars
represent the formal uncertainty in the value of $\alphadm$ from the
power law fit.  The dashed gray lines show the upper and lower limits
to the mass-to-light ratio that we consider reasonable based on the
combination of the stellar population models and the kinematics.  Note
that for small values of \mslk\, the dark matter density profile is
slightly steeper than the total density profile.  This unusual effect
is caused by the steep increase in the \hi\ rotation curve at $r >
60\arcsec$.
\label{mlalpha}}
\end{figure}

We now use Equation \ref{vhaloeq} to obtain the rotation curve due to
the dark halo.  Under the assumption that the density profile can be
described with a power law, $\rhodm \propto r^{-\alphadm}$, we perform
a linear fit to determine $\alphadm$ as a function of \mslk.  The fit
extends out to a radius of 105\arcsec, and we ignore points that have
imaginary halo rotation velocities.  A power law provides a good fit
to the halo rotation curve for any mass-to-light ratio.  The results
of these fits are plotted in Figure \ref{mlalpha}.  For $\mslk > 0.19
\mlk$, $\alphadm < 0$ and the density of the dark matter halo is
\emph{increasing} with radius.  Because such a dark matter
configuration is probably unphysical, we consider 0.19 \mlk\ to be a
firm upper limit to the stellar disk mass-to-light ratio, with the
corresponding \emph{lower limit} to $\alphadm$ of 0.  The dark matter
density profile for this maximal disk is

\begin{equation}
\rhodm = 0.1 \left(\frac{r}{1\mbox{ pc}}\right)^{-0.01 \pm 0.13} 
M_{\odot}\mbox{ pc}^{-3}.
\label{maxdiskdensityprof}
\end{equation}

\noindent
As we argued in \S \ref{popsynth}, the only way that the stellar
mass-to-light ratio can be lower than this value is if the galaxy
contains a young starburst, so Equation \ref{maxdiskdensityprof}
represents the most likely shape for the dark matter halo.  Note that
even though the kinematic value of \mslk\ we derive is rather low,
there are two effects that we have not accounted for that tend to
raise it: the finite thickness of the stellar disk (\S
\ref{thindisk}), and asymmetric drift (\S \ref{asymdrift}).  Including
these effects raises the maximum disk \mslk\ close to the range that
is predicted from the photometry.

The slope of the total density profile of the galaxy represents the
absolute \emph{upper limit} for the slope of the dark matter density
profile, so $\alphadm \le 0.27 \pm 0.09$.  In practice, because the
galaxy contains stars and gas, the upper limit must be lower.  If
NGC~2976 is not undergoing a strong and very young starburst, its
stellar mass-to-light ratio must be at least 0.10~\mlk.

Therefore, we conclude that the dark matter density profile is
bracketed by $\rhodm \propto r^{-0.17 \pm 0.09}$ and $\rhodm \propto
r^{0}$ (see Figure \ref{mlalpha}).  Due to the extremely low value of
the maximal disk mass-to-light ratio, the galaxy must contain an
essentially maximal disk.  We adopt the $\mslk = 0.19 \mlk$ disk and
$\alphadm = 0.01$ halo, which are shown in Figure \ref{diskfig}b, as
our preferred solution for the rest of the paper.  This disk dominates
the gravitational potential of the galaxy out to a radius of
35\arcsec\ (550 pc).  The total mass of NGC~2976 out to the edge of
the observed velocity field at 2.2 kpc is $3.5 \times 10^{9}
M_{\odot}$, of which 5~\% is contributed by the gas, and up to 14~\%
(for $\mslk = 0.19 \mlk$) is contributed by the stars.

\section{DISCUSSION}
\label{discussion}

\subsection{Comparison to Cold Dark Matter Simulations}

In the previous section, we derived the dark matter density profile of
NGC~2976, and determined that it cannot have a slope steeper than
$\alphadm = \alphatot = 0.27 \pm 0.09$.  Even in this minimum
disk case an NFW halo in NGC~2976 is ruled out.

\subsubsection{Does the Density Profile of NGC~2976 Conflict With CDM?}

The shallow central density profile of NGC~2976 does not necessarily
imply a problem with CDM.  It is also possible that the simulations
and the observations may not be directly comparable, or that the
simulations may not incorporate all of the relevant physics.  Our
observations have only probed the very inner portion of the galaxy's
potential, whereas the numerical simulations are better at revealing
the density structure at large radii.  The highest-resolution
simulations can reach radii as small as 0.5~\% of the virial radius
\citep{power03}.  An NFW halo comparable in size to NGC~2976 would
have a virial radius of $\sim80$ kpc, so a simulation resolution
element would be 400 pc in the best case.  There would then be $\sim5$
resolution elements within the observed region of the galaxy, which
might not be enough to accurately determine the slope of the density
profile over those radii.  It is therefore plausible that
higher-resolution simulations could help to resolve the apparent
conflict between the observational and theoretical results.  It is
also worth noting that none of the CDM simulations reported in the
literature to date have explored galaxies as small as NGC~2976.
Although the simulated density profiles appear to be independent of
mass, simulated dwarf galaxies could conceivably have different
density profiles than the large galaxies and galaxy clusters that have
thus far been studied.  Finally, we point out that the current dataset
just reaches what appears to be the peak of the rotation curve; it
would be extremely interesting to trace the rotation curve farther out
as it presumably flattens and turns over.  We are in the process of
using recently-obtained VLA \hi\ observations to carry out this study.

Beyond numerical effects, though, there are more important reasons to
suspect that the simulations may not correspond well to the
observations.  One potentially significant problem with current
simulations is that they neglect the effects of the baryons on the
dark matter halo.  As we have shown, the central region of NGC~2976 is
dominated by the stellar disk.  It is possible that the formation of a
massive disk at the center of a cuspy spherical halo could destroy the
central cusp \citep[][although see \citet{gz02}]{wk02}.

An additional possibility for accounting for the very shallow central
density profile within the context of CDM is suggested by the recent
work of \citet{stoehr02} and \citet{hayashi03}.  These authors find in
their simulations that dark matter satellite halos orbiting in the
potential of a more massive neighbor are subject to tidal stripping.
The stripped satellites end up with density profiles that are much
shallower than their original NFW profiles.  If NGC~2976 can be
identified with one of the most massive few dark matter satellites of
M~81, this mechanism provides a natural way to explain its nearly
constant-density dark matter halo without modifying the CDM model.

We conclude that the solution to the density profile problem does not
currently require fundamental changes to CDM.  There are a number of
simpler explanations that may remedy the discrepancy between
observations and simulations.  More complete simulations can help to
clarify the situation, as can more carefully targeted high-resolution
observations (for example, studies of a few isolated galaxies could
confirm or refute the possibility that density profiles are being
modified by tidal stripping).

\subsubsection{NFW and Pseudoisothermal Fits for NGC~2976}
\label{functions}

Up to this point, we have used power law fits to describe the rotation
curve and density profile, giving us a straightforward measurement of
the central slope\footnote{Note that in general a power law is not a
good representation of the expected CDM density profile form, which
has a logarithmic slope that varies from $\sim-1$ to $-3$.  Our
measurements, however, are all within the characteristic radius of the
halo of NGC~2976, where the density profile predicted by CDM is close
to a power law.}.  This method has two advantages over the traditional
approach of fitting the rotation curve with various observationally or
theoretically motivated functional forms to see which one best matches
the data.  First, it is model-independent.  Second, some functional
forms (NFW, for example) require that the data cover a certain range
of radii in order to constrain the fit parameters.  An NFW rotation
curve reaches its maximum at $2.16 r_{s}$ and then turns over.  If the
velocity data do not extend beyond the turnover radius, the scale
radius (and hence the concentration parameter) of the halo cannot be
reliably measured.

Although we argue that the power-law approach may be more useful, we
recognize that performing NFW and pseudoisothermal fits to our data
will facilitate comparisons to previous work.  Accordingly, we have
used the {\sc rotcurshape} routine (Appendix \ref{rotcurshape}) to
attempt to find best-fitting parameters for the velocity field of
NGC~2976, assuming each of those functional forms for the rotation
curve.  An isothermal halo with a constant-density core provides a
reasonable fit to the data, with a core radius of 67\arcsec\ (1.12
kpc) and an asymptotic velocity of 130 \kms.  This fit is comparable
in quality to the power law fit.  For an NFW rotation curve, {\sc
rotcurshape} cannot obtain a satisfactory fit for any value of the
concentration.  We also attempted to fit the NFW form just to the
rotation curve (not the full velocity field) with various nonlinear
least-squares techniques.  Because we know that the rotation curve of
NGC~2976 is shallower than an NFW rotation curve, we fixed the
concentration parameter at an artificially low value ($c = 9.2$,
$\sim2\sigma$ lower than expected; \citet{bullock01}) for these fits,
and only solved for $v_{200}$ and $r_{200}$.  We found that neither
$v_{200}$ nor $r_{200}$ are significantly constrained by the rotation
curve of NGC~2976.  The best NFW fits have a reduced $\chi^{2}$ value
of 6.2 (compare to a reduced $\chi^{2}$ of 1.3 for a power law fit),
and the NFW rotation curve only passes within $1\sigma$ (combined
statistical and systematic uncertainties) of 2 out of the 27 points in
the rotation curve.  The remainder of the fitted points are up to
$4.1\sigma$ away from the data points, showing that an NFW rotation
curve is very strongly excluded for this galaxy.  Note that both the
pseudoisothermal and NFW fits described here were performed on the
total mass distribution of the galaxy, not just the contribution from
the dark matter halo.  Removing the stellar and gas disk velocities
first would make the NFW fit worse.

Although our velocity field does not extend beyond the turnover of the
rotation curve and NFW fits to the rotation curve are unconstrained,
there is another way to estimate the NFW concentration parameter, and
the effective concentration parameters for other dark halo models from
the data.  \citet{abw02} defined the parameters $R_{V/2}$ (the radius
at which the rotation curve has risen to half of its maximum value)
and $\Delta_{V/2}$ (the mean density within $R_{V/2}$, in units of the
critical density) in order to make it easier to compare rotation curve
observations with theoretical predictions.  For the minimum disk case
in NGC~2976, $V_{max} = 86$~\kms, $R_{V/2} = 768$~pc and $\Delta_{V/2}
= 1.3 \times 10^{6}$.  Using the formulae given by \citet*{abw02}, we
calculate concentrations of 18.5, 4.1, 51.5, and 225.4 for an NFW
profile, a Moore profile, a Burkert profile, and an isothermal+core
profile, respectively.  For our preferred solution, after accounting
for the stellar and gas disks, the dark matter halo parameters are
$V_{max} = 74$~\kms, $R_{V/2} = 902$~pc and $\Delta_{V/2} = 7.0 \times
10^{5}$, reducing the concentrations to 14.5, 3.1, 40.8, and 165.0.

The \citet*{abw02} analysis is designed to study the \emph{value} of
the central density of the dark matter halo (which is also a potential
point of disagreement between observations and simulations).  With or
without accounting for the baryonic contribution to the rotation
curve, the central density of the dark matter halo of NGC~2976
(parameterized by Alam et al.'s definition of $\Delta_{V/2}$) is
consistent with $\Lambda$CDM simulations, even though the shape of the
density profile is not.

\subsection{Comparison to NGC~4605}
\label{4605comparison}

NFW suggested, and most subsequent authors have agreed, that relaxed
CDM halos should all have the same shape independent of mass or merger
history\footnote{Provided that they have not recently undergone a
major merger.  There is no kinematic or photometric evidence to
suggest that either of the galaxies discussed here was recently
involved in a merger.}.  NGC~2976 is superficially rather similar to
the first galaxy we studied, NGC~4605, so it is reasonable to compare
the two.  Our observations of NGC~4605 showed that its dark matter
halo has a density profile with $\alphadm = 0.65$ \citep{us}.  At
first glance, this result does not appear to seriously conflict with
our findings for NGC~2976.  However, the NGC~4605 density profile was
for a maximal disk, and therefore represents a \emph{lower limit} on
$\alphadm$.  We argued that the maximum disk solution was the most
likely for NGC~4605 because the mass-to-light ratio could not
realistically be much smaller than its maximum value of 0.22 \mlk\ in
that galaxy, and because it leads to a simpler density structure for
the halo (a single power law rather than two).

For NGC~2976, by contrast, we set an \emph{upper limit} of
$\alphadm = 0.27$ for the minimum disk case, and we prefer lower
values of $\alphadm$ because a minimum disk is not physically
realistic.  For solutions in the range that we believe is reasonable
($0 \le \alphadm \le 0.17$), the dark matter density profile slope
disagrees with that of NGC~4605 by up to $5\sigma$, even though the
disks of these two galaxies are quite similar.  Although we have only
examined two galaxies so far, their incompatible dark matter
density profiles suggest that the cosmic scatter in halo properties
may be large.

\subsection{Are All Dwarf Galaxies Dark-Matter Dominated?}
\label{dmdominatedsec}

It is generally assumed that, with the possible exception of tidal
dwarfs \citep{bh92}, all dwarf galaxies are dynamically dominated by
dark matter \citep{cf88,cb89,jc90,mcr94}.  While this assumption is
likely true for the outer parts of dwarfs (radii larger than $\sim2$
times the disk scale length), the observational evidence is more
ambiguous close to their centers.  One of the main sources of this
problem is that dwarf rotation curves are traditionally observed in
\hi, with angular resolution as low as 30\arcsec.  The rotation curves
therefore often contain only two or three data points at radii where
the stellar and gas disks are dynamically important.  To make matters
worse, these inner data points are the most likely to be affected by
beam smearing and other systematic problems.  We suggest that higher
resolution observations of dwarf galaxies may show that their central
regions are often dominated by luminous material.

In the case of NGC~2976, the baryonic mass dominates the central 220
pc of the galaxy even for the lower limit to \mslk\ of 0.10 \mlk.  For
our preferred solution of a maximal disk ($\mslk = 0.19 \mlk$), the
disk dominates out to a radius of 550 pc.  Consequently, the stellar
disk has a significant impact on the derived density profile of the
dark matter halo: slopes ranging from $\alphadm = 0.29$ to
$\alphadm = -0.13$ are possible depending on the choice of \mslk\
(see Figure \ref{mlalpha}).

That stars contribute to the dynamics of a dwarf galaxy is not unique
to NGC~2976; similar conclusions were reached for NGC~1560 by
\citet{broeils92}, for NGC~5585 by \citet{blo99}, and for NGC~4605 by
\citet{us}.  In addition, this result is also in agreement with the
work of, e.g., \citet[hereafter PSS]{pss96}, who showed that the
fraction of dark mass in spiral galaxies is a strong inverse function
of luminosity.  PSS found that in galaxies with luminosities
comparable to NGC~2976 ($M_{I} = -18.5$), dark matter can be detected
gravitationally beginning at $10-15 \%$ of the optical radius (which
is located at 2.8 kpc for NGC~2976), or about 350 pc.  This is
entirely consistent with our mass modeling (see Figure
\ref{diskfig}b).  The average $M_{I} = -18.5$ rotation curve
constructed by PSS has dark matter dominating the rotation curve at
radii beyond $0.2 R_{opt}$ (560 pc), also consistent with our
preferred solution.  Thus, even though it may seem counterintuitive,
the PSS results support our conclusion that luminous matter is
sometimes an important contributor to the inner rotation curves of
dwarf galaxies.

\subsection{Are the Kinematics of NGC~2976 Affected By M~81?}
\label{interactionsec}

NGC~2976 does not appear to be participating in the dramatic tidal
interaction currently taking place between M~81, M~82, and NGC~3077
\citep{yhl94}; however, it has likely interacted with M~81 in the
past.  \citet{ads81} discovered a faint \hi\ streamer stretching from
M~81 to NGC~2976.  \citet{boyce01} used HIJASS data to show that this
gas comprises a single tidal bridge that smoothly connects the two
galaxies (see their Figure 2a).  The bridge contains somewhat more
\hi\ than NGC~2976 itself ($2.1 \times 10^{8}$ M$_{\odot}$ and $1.5
\times 10^{8}$ M$_{\odot}$, respectively).  Unfortunately, the HIJASS
observations lack the angular resolution to see the details of the
connection between the bridge and NGC~2976, and the presence of
Galactic \hi\ further complicates the situation.  \citet{yhl00}
suggested that the bridge is a remnant of an interaction that took
place only between M~81 and NGC~2976 before the current
M~81/M~82/NGC~3077 event.  Nevertheless, the optical galaxy (Figure
\ref{colorimage}) and the inner \hi\ disk \citep{si02a,si02b} both
appear regular, symmetric, and undisturbed.  Assuming that M~81 has a
total mass of $\sim10^{12} M_{\odot}$ \citep{k02}, its tidal field
only becomes comparable to the gravity of NGC~2976 (at a radius of 2
kpc) if the galaxies approach within 20 kpc of each other.  Since M~81
is currently at a projected distance of 79 kpc, the present-day
kinematics of NGC~2976 are probably unaffected by the interaction.

\subsection{Possible Origins of Noncircular Motions}
\label{noncirc}

In Figure \ref{velfield}a it is clear that the velocity field of
NGC~2976 is distorted compared to a purely rotating disk.  The
velocity gradient near the center of the galaxy is not directed along
the photometric major axis, but is offset by up to $\sim40\degr$ (see
Figure \ref{ringfitfig}).  This twisting of the isovelocity contours
means that the kinematics of NGC~2976 cannot be described by the
simplest model: a constant PA and only rotational motions.  We have
shown that the velocity field \emph{can} be adequately described by
adding radial motions in the plane of the galaxy to the model.  If
there are systematic trends remaining after this model has been
subtracted from the data, they are only present at the level of a few
\kms\ (see Figure \ref{resids}).  However, a purely rotational
velocity field with a kinematic PA that declines monotonically from
$\sim6\degr$ near the center of the galaxy to $-37\degr$ at a radius
of 90\arcsec, and remains constant at $-37\degr$ for larger radii can
also fit the data. This model is the one produced by \rotcur\ if the
kinematic PA is left as a free parameter (see Appendix \ref{rotcur}).
The total density profile obtained under these assumptions is $\rhotot
\propto r^{-0.56}$.  Since the photometric PA of the galaxy is quite
stable, varying only a few degrees from its average value beyond a
radius of 30\arcsec\ (at smaller radii, local bright spots dominate
the isophotal fitting), this model requires a physical mechanism that
could cause the behavior of the photometric and kinematic PAs to
deviate strongly from one another.  It is unclear what such a
mechanism could be, and why it would make the kinematic PA change so
rapidly.  Because this model lacks an observational motivation, while
radial (or other noncircular) motions are expected to occur naturally
for a variety of reasons (see below), we prefer the radial motion
interpretation of the velocity field.

There are a number of possible sources of the radial motions.  The
galaxy could contain a stellar bar, although there is no sign of a bar
in any of our images, even at 2.2 $\mu$m.  Further evidence against
the presence of a bar is the lack of measurable higher order terms in
our harmonic decomposition of the velocity field.  The velocity field
of a barred galaxy should contain a nonzero $\sin{3\theta}$ term
\citep{wong00}.  An alternative to a bar is the possibility that the
dark halo is triaxial rather than spherical, as we have assumed.  It
is expected that CDM halos should be at least moderately triaxial
\citep{dc91,w92,cl96}, and the potential of a triaxial halo is
certainly not axisymmetric, so the velocity field of a galaxy embedded
in a triaxial halo would exhibit noncircular motions.  However, since
the details of such a velocity field have not yet been simulated, we
cannot compare our results to theoretical predictions.  Future
simulations of the kinematics of a gaseous disk within a triaxial halo
would be quite interesting.  Other potential causes of the radial
motions include a disk that has nonzero ellipticity, and outflows
associated with star formation.

\section{SYSTEMATICS}
\label{systematics}

In this section, we study in detail the systematic uncertainties in
our analysis, and also some systematic problems that afflict rotation
curve studies in general.  We emphasize that systematic effects are
the dominant source of uncertainties in our analysis.  Some of the
details contained in this section are therefore crucial to
understanding the reliability of our conclusions. The general reader
may wish to use the summary in the following paragraph and the
subsection headings to select the portions in which he or she is
interested.

We begin in \S \ref{sysintro} by mentioning the importance of
considering systematic problems, and our efforts to account for these
problems in the design of our survey.  Section \ref{rcsystematics}
continues with a description of our tests for systematic errors caused
by the rotation curve fitting.  In \S \ref{velcompare} we demonstrate
that the \ha\ and CO velocity fields of NGC~2976 are consistent with
each other, not just globally, but on a point-to-point basis.  In \S
\ref{longslit} we use our velocity field to simulate longslit
observations of NGC~2976, and compare the derived longslit density
profiles to the one we extract from the two-dimensional velocity data.
Section \ref{wrongcenter} examines the problem of offsets between the
kinematic center of a galaxy and the position of the slit during
spectroscopic observations, and \S \ref{bars} briefly discusses the
difficulties that barred galaxies present for density profile studies.

\subsection{The Problem of Systematics}
\label{sysintro}

It is well-known, although not often discussed, that there are a
number of serious systematic uncertainties that can cause an observed
rotation curve (and the associated density profile) to differ
significantly from the true one.  Worse, nearly all of these effects
work in the same direction to cause density profiles to appear
systematically shallower than they actually are.  Fortunately, the
most severe of these problems can be minimized or avoided by using
two-dimensional velocity fields and by making velocity measurements at
very high precision \citep{blo99,vdbs01,swaters02,us}.  SMVB model
several of these effects in detail and determine how severely
observational results may be affected in various situations.  One of
the key systematics to investigate is the location of the dynamical
center of the galaxy with respect to the assumed center.  Other
systematic errors that might potentially cause problems for rotation
curve studies include extinction (for \ha\ observations), beam
smearing (for \hi\ observations), noncircular motions (which could be
caused by a bar, an intrinsically elliptical disk, a triaxial halo, or
outflows associated with vigorous star formation), incorrect galaxy
centers, PAs, inclinations, and systemic velocities, asymmetric drift,
and using observations made at low velocity resolution to study
galaxies with small rotation amplitudes.  We designed our study so as
to avoid some of these problems, and to be able to account for the
others, as described in the introduction; a few key points are
mentioned below.  We know that extinction does not affect our results
because our CO and \ha\ velocity fields agree near the center of the
galaxy, where extinction should be most important.  Since we have
two-dimensional information, we can explicitly account for radial
motions, as discussed in \S \ref{darkmatter}, \S \ref{rcsystematics},
and Appendix \ref{algorithms}.  Our velocity resolution is $\sim10$
times smaller than the maximum rotation velocity of NGC~2976 (and the
precision of our velocity measurements is another order of magnitude
smaller), so we are unlikely to be missing significant features in the
velocity field due to insufficient resolution.

\subsection{Rotation Curve Fitting Systematics}
\label{rcsystematics}

Due to the high precision of our velocity measurements, the
statistical error bars on both the rotation curve and the
radial velocity curve are negligible (less than 1 \kms\ everywhere).
Therefore, the limiting factors on the accuracy of the rotation curve
are the systematic uncertainties associated with our fit, which is a
normal state of affairs for rotation curve and density profile
observations.

\subsubsection{Algorithmic Differences}
\label{algodiff}

The most straightforward check for systematic problems is to compare
the rotation curves produced by different algorithms.  Recall that
neither \rotcur\ nor \rotcurshape\ allow the user to fit for the
$\sin\theta$ term (radial motions) in Eq. \ref{ringfitvmodel}, and
\ringfit\ and \rotcurshape\ both require that the PA, inclination, and
center position not vary with radius.  Using a set of input parameters
that are compatible with all three algorithms (PA$=-37\degr$,
$i=61.5\degr$, $\cos{\theta}$ weighting, and no radial motions) and
considering radii less than 105\arcsec, the algorithms all produce
essentially identical results.  We conclude that none of the
assumptions that are built in to the fitting algorithms affect the
results.

The only significant difference that appears between the algorithms
stems from the inclusion of radial velocities in the fit.  Earlier, we
noted that it is obvious from inspection of the velocity field (Figure
\ref{velfield}) that noncircular motions are present in NGC~2976: for
example, the velocity fits in individual rings for $r<40\arcsec$ show
that the observed velocity maximum is systematically offset from the
photometric major axis (Figure \ref{ringfitfig}).  Neglecting the
$\sin\theta$ term and fitting only for rotation increases the exponent
of the density profile from $\alphatot=0.27$ to $\alphatot=0.42$ (for
$0<r<105\arcsec$).

If radial motions are ignored, however, an accurate description of the
velocity field requires that that kinematic PA changes with radius.
When either the kinematic PA or the inclination varies with radius,
\rotcur\ yields $\alphatot \approx 0.56$.  If we allow both parameters
to change with radius, tying the inclination to the photometric axis
ratio and fitting for the kinematic PA, we obtain total density
exponents as high as $\alphatot=0.7$. Thus, by exchanging radial
motions for geometric degrees of freedom it is possible to push
$\alphatot$ to higher values. However, we regard these models as
contrived and lacking physical motivation, and therefore less
appealing than simply including radial motions.  

\subsubsection{Uncertainty in Center Position}
\label{dcenter}

Assessing the uncertainties on the rotation velocities requires that
we first know the uncertainties on each of the parameters that are
used to calculate the rotation velocities: the center, PA, inclination,
and systemic velocity.  We begin by considering the center position of
NGC~2976.

The galaxy nucleus is located within 3\arcsec\ of previously published
estimates of the galaxy's position.  The astrometric precision on the
photometric location of the nucleus is 0\farcs2.  However, the
velocity field is only aligned with the images to about 1\arcsec, and
the resolution of the velocity field is 4\arcsec, which limits the
degree to which we can verify that the nucleus and the kinematic
center of the galaxy coincide.  To determine the position and
uncertainty of the kinematic center, we used a bootstrap resampling
technique.  By running \ringfit\ on 200 bootstrap samples of the
velocity field, we measured a kinematic center of \ad =
(09\hr47\min15.5\sec, 67\degr55\min00.2\sec), with an uncertainty of
2\arcsec\ in both $\alpha$ and $\delta$.  Thus, there is no evidence
for an offset between the kinematic and photometric centers of
NGC~2976.

\subsubsection{Uncertainty in Position Angle}
\label{dpa}

We used the same bootstrap method to measure the kinematic
PA\footnote{The kinematic PA is distinct from the photometric PA in
that it is defined as the angle between north and the \emph{receding
side} of the galaxy's major axis, so that it has a range of
$-180\degr$ to $180\degr$ (where positive angles are east of north).}
and its uncertainty.  The kinematic PA is well-determined at PA$_{kin}
= -36\degr \pm 5\degr$ and agrees with the photometric PA.

\subsubsection{Uncertainty in Inclination Angle}
\label{dinc}

Since the photometric inclination of NGC~2976 is relatively high, an
error in the inclination angle does not significantly change the
rotation velocities.  Furthermore, because changing the inclination by
a small amount is approximately equivalent to scaling the rotation
curve by a constant, the exponent of the power law fit should not be
affected.  The reader may recall that if the ellipticity is left as a
free parameter during the surface brightness profile fitting (\S
\ref{sbprofsec}), {\sc ellipse} calculates inclinations that vary
smoothly between 55\degr\ and 77\degr\ across the galaxy.  As
mentioned before, this behavior is not interpreted as an actual change
of the inclination angle with radius.  Nevertheless, if we force
\rotcur\ to use this function for the inclination, the density profile
slope for the total mass distribution increases to
$\alphatot\approx0.56$, as discussed in \S\ref{algodiff}.

\subsubsection{Uncertainty in Systemic Velocity}
\label{dvsys}

If the systemic velocities are left as a free parameter in the
velocity field fits, they have a scatter of 1.8 \kms\ from ring to
ring.  It is possible that these variations are real, although they
are only marginally significant when the systematic uncertainties are
taken into account.  The alternative approach is to fix $v_{sys}$ at
its average value and then fit again with only $v_{rot}$ and $v_{rad}$
as free parameters.  The results of the fit with $v_{sys}$ fixed are
nearly identical to the previous results.  None of the radial or
rotational velocities are changed by more than $1\sigma$, the density
profile exponent for the total mass distribution increases by less
than $1\sigma$ (to $\alphatot = 0.34 \pm 0.09$), and the maximum
allowed mass-to-light ratio increases to 0.24 \mlk.

\subsubsection{Uncertainties in Rotation Velocities and Radial Velocities}
\label{dvrot}

Using the measured uncertainties in the center position and PA, and
assuming an uncertainty of 3\degr\ for the inclination angle, we
calculated the resulting uncertainties on the rotation velocities and
the radial velocities with a Monte Carlo technique.  We generated 1000
random centers, PAs, and inclinations, assuming a Gaussian
distribution for each of the parameters, and ran \ringfit\ with each
set of parameters.  The standard deviation of the 1000 rotation
velocities in each ring was defined to be the systematic error of that
rotation velocity, and the systematic errors in the radial velocities
and systemic velocities were calculated in the same way.  The
systematic errors on the rotation curve range from 2.1 \kms\ to 5.5
\kms, as listed in Table \ref{data}.  Power law fits to the 1000 Monte
Carlo rotation curves yield a mean slope of the total density profile
of $\alphatot = 0.27 \pm 0.09$.

\subsubsection{Asymmetric Drift Correction}
\label{asymdrift}

We have also calculated the asymmetric drift correction to the
rotation curve, as defined by, e.g., \citet{ccf00}.  We derived the
velocity dispersion $\sigma$ as a function of radius from the \ha\
data, and the surface density $\Sigma$ by adding the \hi\ and \htwo\
column densities.  We then fit polynomials to $\sigma(r)$ and
$\Sigma(r)$ and calculated the derivatives $d\sigma / d\ln{r}$ and
$d\ln{\Sigma} / d\ln{r}$ analytically.  There are significant
uncertainties that factor into this calculation, including 1) we have
not included the ionized gas surface density (although its
contribution is expected to be small), 2) the \htwo\ surface density
is uncertain due to our imprecise knowledge of the CO-\htwo\
conversion factor, 3) our velocity field extends to radii that are
smaller than the resolution of the \hi\ data, so that the calculated
asymmetric drift may be incorrect for the inner few points of the
rotation curve where the correction is most important, and 4) some of
the \ha\ velocity dispersion is probably due to flows associated with
star formation.  Because we view the derived corrections as rather
uncertain, we have not corrected the observed rotation curve for
asymmetric drift.  However, the corrections are listed in Table
\ref{data} if the reader wishes to apply them.  Their effects are to
increase the maximum disk mass-to-light ratio, and to make the
rotation curve slightly more linear, but the overall conclusions of
the paper do not change.

\subsubsection{Conclusions From Analysis of Rotation Curve Systematics}
\label{systematicconclusions}

We have shown that the only ways to significantly change the derived
slope of the density profile of NGC~2976 are to 1) assume the stellar
mass-to-light ratio is zero, 2) ignore the radial component of the
velocity field, or 3) allow the kinematic PA and/or inclination to
change with radius.  Assuming that the observed velocities are due
entirely to rotation raises $\alphatot$ by $\sim0.15$, and allowing
the PA and inclination to change with radius raises $\alphatot$ by up
to an additional $\sim0.25$.  Accounting for the contribution of the
maximum stellar disk, however, limits the dark matter density profile
exponent to $0.26 \le \alphadm \le 0.4$.

Because inspection of the velocity field and the fits clearly reveals
the presence of radial motions, neglecting the radial component is not
justified.  Changes in the PA with radius are not supported by the
photometry, and changes in the inclination with radius are difficult
to understand physically.  Therefore, we argue that these solutions,
despite being mathematically viable, are contrived and not motivated
by the data.

We conclude that the galaxy contains substantial radial motions, and
that the density profile results are not significantly affected by the
most obvious sources of systematic errors.  We caution that the
robustness against systematics that we find is specific to this
dataset, and may not be true in general.  Because the rotation curve
of NGC~2976 increases with radius so slowly, errors in any of the
geometric parameters of the galaxy are diminished in importance.  A
galaxy with a more rapidly rising rotation curve would probably be
more severely affected.  Assuming that the radial motions provide no
support, the dark matter density profile slope is in the range $0 \le
\alphadm \le 0.27$, with a $2\sigma$ upper limit of $\alphadm \le
0.45$, where systematic errors have been included in the uncertainty
on $\alphadm$.  NGC~2976 thus violates the prediction of universal
central density cusps by CDM simulations.

\subsection{Comparing Velocities Derived From Different Tracers}
\label{velcompare}

Some recent studies in the literature have shown that, beam smearing
questions aside, there do not appear to be systematic offsets between
\hi\ and \ha\ rotation velocities \citep[e.g.,][]{mrdb01,m02}.  With a
handful of exceptions, though, these studies employed longslit \ha\
data, so the comparisons essentially took place only along the major
axis.  In addition, the spatial and velocity resolution of the \hi\
and \ha\ data were often quite different.

\begin{figure}
\figurenum{12}
\epsscale{0.5}
\plotone{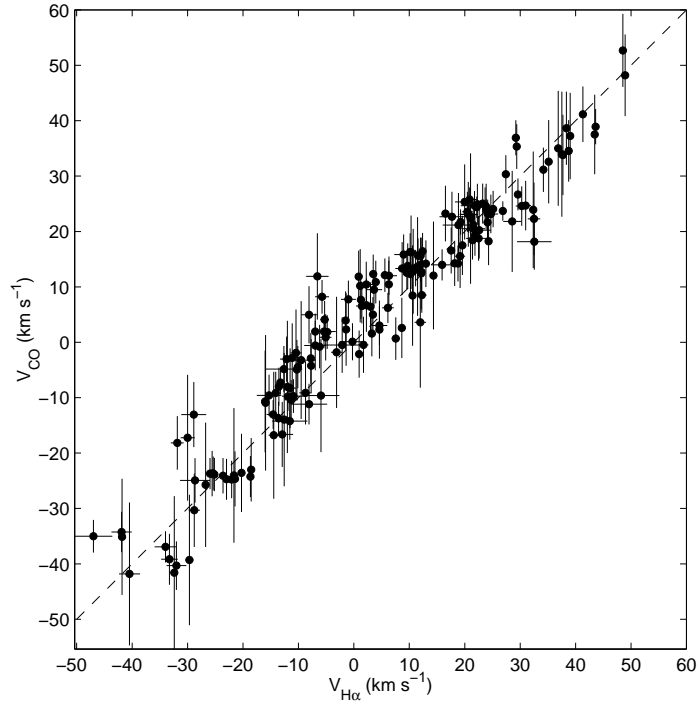}
\caption{Point-by-point comparison of \ha\ and CO velocities.
Although the line \vha\ = \vco\ provides a very good description of
the data, with remarkably small scatter, there are still small
systematic trends visible near the center of the galaxy and at
positive velocities.
\label{havsco}}
\end{figure}

In this paper, we have presented for the first time the data necessary
for a two-dimensional comparison across a dwarf galaxy of the CO and
\ha\ velocity fields.  The angular resolution of the two datasets is
similar (6\arcsec\ and 4\arcsec, respectively), and although the CO
velocity resolution is better by a factor of $\sim6$, the higher
signal-to-noise at \ha\ compensates such that the velocities can be
measured with comparable precision.  We use the following technique to
compare the velocity fields.  At the position of each \ha\ fiber, we
compute a weighted average of the velocities of all of the pixels in
the CO map that fall within the radius of the fiber.  CO pixels that
do not contain any emission are not used in computing the average, and
of course, \ha\ fibers that do not coincide with any molecular
emission are not used either.  This process yields a unique one-to-one
mapping between the two velocity fields.  The rms difference between
\vha\ and \vco\ is 5.3 \kms, with the comparison being made at 173
points.  Similar studies in the Milky Way found that the dispersion
between the velocities of molecular clouds and the associated
\ha-emitting gas was 4-6 \kms\ \citep{ftb82,fdt90}, so much of the
scatter we observe in NGC~2976 may be intrinsic to the process of
\hii\ region formation rather than caused by observational
uncertainties.  We plot the \ha\ velocities against the CO velocities
in Figure \ref{havsco}.  There is a weak systematic trend visible in
the residuals, with \vco\ $>$ \vha\ near the center of the galaxy and
\vco\ $<$ \vha\ on the northwest side of the galaxy.  The amplitude of
this trend is only a few \kms, so it does not affect our rotation
curve.  The origin of the trend is not clear, but we suggest that it
could be a result of the spatial distribution of the gas.  For
example, where the ionized gas is largely in front of the molecular
clouds, the expansion of \hii\ regions away from nearby molecular
clouds would make \vha\ $>$ \vco.  This effect should appear
preferentially where \ha\ emission is bright.  Conversely, where the
molecular clouds are in front, one would expect that \vha\ $<$ \vco.
These areas should have faint \ha\ emission due to extinction within
the molecular clouds.  The \ha\ distribution in NGC~2976 appears to be
qualitatively consistent with this interpretation; the \ha\ is
brighter in the northwest, where the \ha\ velocities are larger, and
there is an \ha\ hole to the southeast, where some of the CO
velocities are higher.  In any case, we conclude that the \ha\ and CO
velocity fields agree, with a scatter of 5.3 \kms, and thus that both
species should be accurate tracers of the gravitational potential of
NGC~2976.

\subsection{Simulated Longslit Observations of NGC~2976}
\label{longslit}

\begin{figure}
\figurenum{13}
\epsscale{0.5}
\plotone{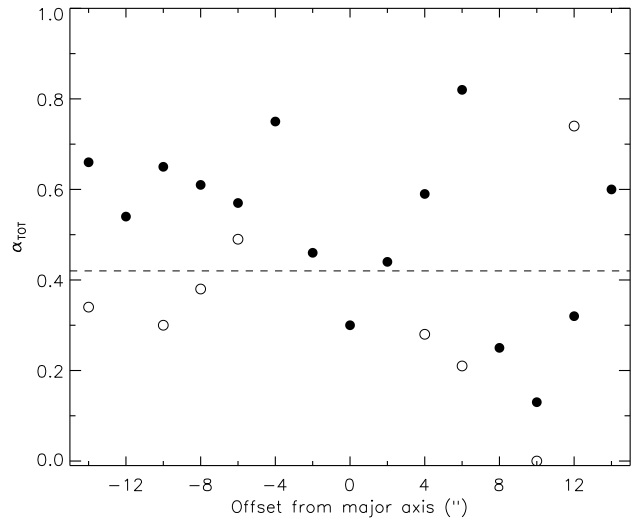}
\caption{Density profile slopes derived from simulated longslit
observations of NGC~2976.  The filled circles indicate the derived
value of the density profile slope $\alphatot$ for each offset from
the major axis.  The dashed line shows the value of $\alphatot$ from
our analysis of the full velocity field (without radial motions, since
they cannot be accounted for in longslit observations).  The open
circles represent the slopes that would have been derived had the
correct (closest to the actual center) folding point been selected for
each of the slits.  For some slits, the value of $\alphatot$ is quite
sensitive to the choice of the folding point.  Note that the open
symbol at (10,0.0) should actually be located at $\alphatot=-0.21$,
except that we do not allow negative values of $\alphatot$ because
they are unphysical.
\label{longslitfig}}
\end{figure}

Our \ha\ dataset is well suited for studying the systematic problems
associated with deriving rotation curves from longslit spectroscopy.
It is straightforward to recreate what would be seen by an observer
taking longslit spectra of NGC~2976.  We begin by selecting all of the
fibers within 1\arcsec\ of a given cut parallel to the major axis of
the galaxy.  This creates an unevenly-sampled rotation curve, which we
smooth by averaging the points into 4\arcsec-wide bins.  We then
proceed exactly as we would if we had obtained these rotation
velocities from a longslit spectrograph.  We find the center of the
rotation curve by folding it about various points to determine the
position of maximum symmetry.  Three criteria are used to judge the
degree of symmetry: the correlation coefficient of the two sides, the
rms difference in velocity between points at the same radius on
opposite sides, and the appearance of the rotation curve.  These
criteria are combined in a necessarily somewhat subjective manner, but
since we know the true center in this case from our two-dimensional
velocity field, we have verified that the chosen center never differs
from the actual one by more than 12\arcsec\ (200 pc).  We fold the
rotation curve about the chosen center and average the two sides
together, weighting each point by the inverse square of its
uncertainty.  Finally, we fit a power law to the resulting rotation
curve, ignoring any points near the center that have negative rotation
velocities.  We repeat this process with offsets from the major axis
of up to 14\arcsec\ (230 pc).  The indices, $\alphatot$, of the
power law fits in density for each rotation curve are displayed in
Figure \ref{longslitfig}.

The naive expectation from this experiment is that slits placed off of
the major axis will make the density profile appear to be shallower
than it actually is, and that this effect should become more severe
with increasing distance from the major axis.  The actual results do
not show this trend very clearly.  The positive slit offsets
(corresponding to the northeast side of the galaxy) appear to agree
with the expected behavior; for large slit offsets the slopes are on
average shallower than the value that should be derived ($\alphatot =
0.42$, since we are neglecting radial motions).  Offsets on the other
side of the galaxy, though, do not follow a systematic trend.  The
derived slopes for negative offsets are similar to the actual slope.
Note that when the rotation curve is folded about the correct central
position (instead of the one that gives the most symmetric
appearance), the slopes are systematically shallower than when other
central positions are used.  We speculate that this systematic effect
is not very strong in NGC~2976 because this galaxy has a relatively
shallow central velocity gradient.  Other galaxies with steeper
rotation curves might be affected more severely.  A possible
explanation for the difference between the two sides of the galaxy is
that the \ha\ distribution is rather asymmetric; to the northeast of
the major axis are a number of bright point sources, while the
emission on the southwest side is faint and diffuse.  In addition to
coherent systematic errors, this exercise shows that attempting to
derive a density profile from a single velocity cut through a galaxy
is also a noisy process.  Depending on the position of the slit, one
could estimate a density profile between $\rhotot \propto r^{-0.13}$
and $\rhotot \propto r^{-0.82}$ for this galaxy.  Only with
observations along many slits, or full two-dimensional velocity data,
can one be confident that the rotation curve and density profile of a
galaxy accurately reflect its gravitational potential.

\subsection{Positioning Errors and Slit Offsets}
\label{wrongcenter}

There are several factors that can play a role in positioning errors.
First is the pointing and guiding of the telescope used to acquire the
data.  \citet*{mrdb01} and \citet{dbb02} report that observations of
LSB galaxies with different telescopes and instruments by independent
observers result in essentially identical rotation curves.  On this
basis, they conclude that pointing errors do not impact their results.
Telescope pointing and guiding are thus unlikely to cause problems for
longslit observations, although they can be an issue for
two-dimensional observations like ours, where the galaxy may not be
visible on the guiding camera while observing it (see \S \ref{hadata}
for our solution to this problem).  Quite independent of the telescope
pointing, though, is the question of whether the center of a given
galaxy is where it is reported to be.  Even when the photometric
center is known accurately, galaxies can have dynamical centers that
differ from the photometric ones by hundreds of parsecs
\citep{puche91,hb95,mg02}.  Because of these two additional problems,
a demonstration that pointing errors are minimal does not suffice to
prove that rotation curves are systematically unaffected by offsets
between their assumed and actual centers.

One example in the literature of a galaxy in which a poorly-determined
center may have caused erroneous conclusions about its density profile
is NGC~2552, often referred to by its alternate name of UGC~4325.  The
rotation curve of this dwarf LSB galaxy has been discussed in a number
of recent papers \citep*{vdbs01,db01a,swaters02,dbbm02,m02,dbb02}.
All of these authors find density profiles with $\alpha \approx 0.3$,
where $\alpha$ is the central slope of the density profile (see \S
\ref{dmlimits} or Equation \ref{alphaeq}).  With six independent
analyses reaching the same conclusion, it would seem that the density
profile of this galaxy is well-determined.  However, closer inspection
reveals a potentially important discrepancy between these studies:
they assume widely varying central positions for the galaxy (see Table
\ref{NGC2552table}).  Two papers \citep*{vdbs01,swaters02} measure the
galaxy's center from their own photometry, while the other four make
no reference to the assumed center.  It is reasonable to suppose that
they used the coordinates given by one of the standard databases, such
as the NASA/IPAC Extragalactic
Database\footnote{http://nedwww.ipac.caltech.edu/} or the SIMBAD
Database\footnote{http://simbad.u-strasbg.fr/}.  These different
positions span a range of 11\arcsec, or 550 pc at the distance of
NGC~2552.

The reason for the uncertainty in the galaxy center is clear from
inspection of a Digitized Sky Survey image: NGC~2552 is a lopsided
galaxy, with a low surface brightness outer disk that is off center
relative to the brighter inner disk.  However, there has recently been
an accurate determination of the actual position of the galaxy;
\citet{bsvdm03} used HST imaging to show that NGC~2552 contains a
nuclear star cluster, and that this nucleus is also the center of the
inner isophotes.  Because galaxies can display offsets between their
nuclei and their dynamical centers, it is possible that the nucleus is
not located at the dynamical center of NGC~2552, but in the absence of
two-dimensional kinematic information it represents the best guess.
As can be seen in Table \ref{NGC2552table}, the previously-used
positions are up to 9\arcsec\ (450 pc) away from the nucleus.  Based
on the results of SMVB and our discussion in Section \ref{longslit},
we suggest that density profiles derived from the longslit
observations cited above could have been significantly over or
underestimated.  \citet*{dbbm02} argue that this is not the case
because the three slit positions they used to observe NGC~2552 (one
through their assumed center, and the other two 5\arcsec\ away on
either side) result in similar density profiles.  However, one of the
three slopes they measure ($\alpha = 0.32,-0.16,0.30$) differs from
the others by $3 \sigma$, showing that slit offsets can cause density
profiles different from the true one to be derived.  Also, if none of
the three slits went through the actual center of the galaxy, then
further observations may be needed to ensure that the measured density
profile is correct.  Given this degree of confusion over the position
of a relatively nearby (d = 10 Mpc) and high-surface brightness (for a
galaxy classified as LSB; $\mu_{R} = 21.6$ mag arcsec$^{-2}$) galaxy,
it is certainly not obvious that the centers of fainter and more
distant galaxies are well-determined in the literature.

\subsection{Barred And Highly Inclined Galaxies}
\label{bars}

Two other common attributes of galaxies that can cause systematic
errors deserve mention here: bars and high inclination angles.  SMVB
already discussed the problems associated with galaxies that are seen
edge-on or nearly so and demonstrated that observations of such
galaxies must be analyzed with extreme care.  Equally problematic,
though, are galaxies that contain bars.  Barred galaxies certainly
have noncircular motions out to the end of the bar, so one-dimensional
velocity data will be systematically affected.  Density profiles of
barred galaxies derived from longslit data are therefore not
trustworthy at radii less than the bar semimajor axis.  SMVB include
five barred galaxies in their sample, and unsurprisingly find that
these objects have shallower central density slopes than their other
targets.  The kinematics of barred galaxies are interesting in their
own right; there are suggestions that the presence of a bar can affect
the evolution of dark matter density cusps \citep{wk02}, and with
two-dimensional velocity fields it is possible to use barred galaxies
to study dark matter density profiles \citep{weiner1,weiner2}.
However, longslit observations of barred galaxies may not be a
reliable way to attack the density profile question.

\section{CONCLUSIONS}
\label{conclusions}

We have used two-dimensional velocity fields, sampled at high spatial
resolution and high spectral resolution in CO and \ha\ to study the
density profile of NGC~2976 and the parameters of its stellar disk and
dark matter halo.  We obtained rotation curves from the
two-dimensional data using tilted-ring models derived with three
independent and complementary algorithms.  Our tilted-ring fitting
shows that there are significant radial (i.e., noncircular) motions in
the inner 20\arcsec\ (300 pc) of the galaxy.  Accounting for these
motions yields a total density profile of $\rhotot \propto r^{-0.27
\pm 0.09}$.  There is a narrow range of possible stellar mass-to-light
ratios for NGC~2976, and the corresponding dark matter halo density
profiles range from $\rhodm \propto r^{-0.17 \pm 0.08}$ to $\rhodm
\propto r^{-0.01 \pm 0.13}$ (constant density).  A key assumption that
we make in the inversion of the rotation curve to obtain the density
profile is that the gravitational and centripetal forces are in
equilibrium (or equivalently, that the radial motions provide no
support).  The density profile obtained by excluding measurements
inside the 20\arcsec\ radius is identical to that computed when
including them, substantiating this assumption.

We found that in our preferred model, the maximum mass-to-light ratio
of the stellar disk of NGC~2976 is $\mslk = 0.09^{+0.15}_{-0.08}
\mlk$.  If $\mslk > 0.19 \mlk$, the dark matter halo has the
unphysical property that its density increases with radius.
Accounting for the thickness of the stellar disk and the asymmetric
drift correction to the rotation curve brings this kinematic value of
\mslk\ into line with photometric estimates.  Comparison with stellar
population synthesis models \citep{worthey94,l99,bdj01} suggests that
the mass-to-light ratio is unlikely to be less than 0.10 \mlk, so the
stellar disk --- and hence the dark matter halo --- are tightly
constrained.  We investigated many of the likely systematic effects on
the rotation curve and found that none of them can bring the density
profile close to $\rhodm \propto r^{-1}$.  We therefore rule out an
NFW or \citet{moore99b} density profile in the center of this galaxy
at high confidence regardless of the stellar contribution.

In addition, we investigated the most extreme models of the galaxy
that are allowed by the data.  Density profile slopes as high as
$\alphadm \sim 0.7$ can be obtained, \emph{but only when all three of
the following are true:} 1) the mass-to-light ratio of the matter in
the disk is zero, 2) the observed velocities are attributed entirely
to rotation, despite the observed radial motions, and 3) the kinematic
PA and inclination both change with radius in the manner described in
\S \ref{noncirc} and \S \ref{systematicconclusions}.  Retaining
requirements 2 and 3, but assuming a maximal stellar disk, reduces
$\alphadm$ to $\lesssim 0.4$.  We consider these models to be quite
unlikely, and inconsistent with the complementary data we have
presented for this galaxy.

We also discussed whether a universal dark matter halo shape is
consistent with our observations.  In a similar study of the slightly
more massive galaxy NGC~4605, \citet{us} found a lower limit to the
dark matter density profile slope of $\alphadm = 0.65$.  Since the
upper limit for NGC~2976 is $\alphadm = 0.27$, the density profiles
of the halos of these galaxies are different from one another.  If the
disk of NGC~4605 is submaximal, or the disk of NGC~2976 is not minimal
(which is likely), the inconsistency becomes more severe.  In
addition, we note that both of these dwarf galaxies are dynamically
dominated by luminous matter at small radii.

Finally, we considered the impact of some of the known systematic
uncertainties that afflict rotation curve studies, following up on the
recent work of SMVB and \citet*{dbbm02}.  We found that
\emph{systematic errors can cause the density profiles inferred from
longslit observations to differ significantly from the true density
profiles}.  We also illustrated the difficulties that can arise in
determining the positions of galaxy centers without adequate
two-dimensional kinematic and photometric information.  These problems
--- as well as the disk-halo degeneracy --- can be largely overcome by
using high-resolution two-dimensional velocity fields, as we have
shown in this paper.

Although previous studies have found that central density cusps cannot
be ruled out in many dwarf and LSB galaxies, we have demonstrated that
a cusp is not present in NGC~2976; the dark matter halo of this galaxy
is nearly constant density out to the edge of the observed \ha\
emission at a radius of 2.2 kpc.

\acknowledgements{This research was supported by NSF grant
AST-9981308.  We thank the referee, Rob Swaters, for suggestions that
improved the paper.  In addition, we would like to thank Di Harmer for
her assistance both with the preparation of our observing proposal and
with using DensePak, and we acknowledge the telescope operating skills
of Gene McDougall and Hillary Mathis.  We also thank Amanda Bosh for
her help with our Lowell observing.  JDS gratefully acknowledges the
invaluable assistance of Peter Teuben in getting \rotcur\ up and
running and then modifying the code to better suit our needs.  This
publication makes use of data products from the Two Micron All Sky
Survey, which is a joint project of the University of Massachusetts
and the Infrared Processing and Analysis Center/California Institute
of Technology, funded by the National Aeronautics and Space
Administration and the National Science Foundation.  This research has
also made use of the NASA/IPAC Extragalactic Database (NED) which is
operated by the Jet Propulsion Laboratory, California Institute of
Technology, under contract with the National Aeronautics and Space
Administration, NASA's Astrophysics Data System Bibliographic
Services, the SIMBAD database, operated at CDS, Strasbourg, France,
and the LEDA database (http://leda.univ-lyon1.fr).  Finally, we would
like to thank Wendy and Liliana for allowing us to cruelly abandon
them in order to observe in such faraway places as Arizona and
northern California.}

\appendix
\section{ROTATION CURVE FITTING ALGORITHMS}
\label{algorithms}

When dealing with longslit kinematic data it is relatively
straightforward to transform the reduced observations into a rotation
curve.  For a full velocity field, the process is more complicated
because it involves converting two-dimensional data to one dimension
while retaining as much of the information as possible.  In this
appendix, we describe the various techniques we use to make this
conversion.

\subsection{Rotcur}
\label{rotcur}
{\sc Rotcur} \citep{b87} is a standard algorithm to fit galaxy
kinematics with a tilted-ring model.  We used the implementation of
\rotcur\ in the NEMO package \citep{teuben}.  {\sc Rotcur} divides
the galaxy into a set of narrow, concentric rings, and in each ring
performs a nonlinear least squares fit to the function

\begin{eqnarray}
\nonumber
\lefteqn{v_{model}(x,y) = } \\
&& v_{0} + v_{rot} \sin{i} \frac{-(x-x_{0}) \sin{PA} + (y-y_{0}) \cos{PA}}
{\sqrt{(x-x_{0})^2 + (y-y_{0})^2/\cos{}^{2}i}},
\label{vmodel}
\end{eqnarray}

\noindent where $v_{0}$ is the systemic velocity, $v_{rot}$ is the
rotation velocity, $i$ is the inclination angle, $PA$ is the angle
between north and the receding side of the galaxy's major axis, and
($x_{0},y_{0}$) is the galaxy's center.  Each ring can thus contain up
to six free parameters (the central position requires two), and
\rotcur\ finds the best fit by minimizing $\sum_{i} (v_{obs,i} -
v_{model,i})^{2}/w_{i}^{2}$, where $w_{i}$ is the weight ascribed to
each point.  We weight each point by the cosine of the angle between
the point and the major axis, automatically deemphasizing points near
the minor axis, so it is not necessary to discard points within some
angle of the minor axis.  {\sc Rotcur}'s most serious weakness
is that it can only model rotational motions.

To create the \rotcur\ rotation curve, we used the best-fit center and
systemic velocity that we determined with \ringfit.  Because the
position angle must be a function of radius if the galaxy is modeled
with purely rotational motions, we first ran \rotcur\ with both the
rotation velocities and the position angle free to vary to determine
$PA(r)$.  We then used this description of the position angle as an
input to \rotcur, and ran it again with only the rotation velocities
as free parameters.  The rotation curve produced in this way is
displayed in Figure \ref{rcfig}b.  We did not allow \rotcur\ to fit
for the inclination angle because it was apparent early on that the
rotation curve of NGC~2976 is essentially solid-body, which means that
$dv_{rot}/dr$ is small.  Therefore, the inclination angle and the
rotation velocities are degenerate in Equation \ref{vmodel}, making the
kinematic inclination angle poorly determined.  We judged that the
inclination angle was unlikely to differ significantly from the
photometric value anyway, so the safest course was to leave the
inclination fixed at 61.5\degr.

\subsection{Ringfit}
\label{ringfit}
In addition to \rotcur\ we constructed tilted-ring models with our own
routine, \ringfit.  The purpose of this exercise was twofold: first,
to compare the results from \rotcur\ with those from a completely
independent program and make sure that the answers agreed, and second,
to fit for radial motions in the plane of the galaxy (inflow or
outflow) instead of just assuming that the observed velocity field was
due only to rotation.  The \ringfit\ fitting function is similar to
Equation \ref{vmodel}, except that we add an extra term to allow for
radial velocities, and we do not fit for the PA, inclination, or the
center.  Thus, we can drop the explicit mention of the PA, $x_{0}$,
and $y_{0}$, and write

\begin{equation}
v_{model} = v_{0} + v_{rot} \sin{i} \cos{\theta} + v_{rad} \sin{i} \sin{\theta},
\label{ringfitvmodel}
\end{equation}

\noindent where $\cos{\theta}$ is equal to the fractional expression
that follows $v_{rot} \sin{i}$ in the second term on the right hand
side of Equation \ref{vmodel}, and the free parameters in each ring
are $v_{0}$, $v_{rot}$, and $v_{rad}$.  The solution is then
determined with a linear least squares fit.  The inclination, PA, and
central position must be specified as inputs, but they can also be
solved for by running {\sc ringfit} with a grid of input parameters
and minimizing $\chi^{2}$.  We have verified that \rotcur\ and
\ringfit\ give indistinguishable results when the same input
parameters and weighting function are used.

\subsection{Rotcurshape}
\label{rotcurshape}

We also employed the NEMO routine \rotcurshape, which is closely
related to (and based on) \rotcur.  {\sc Rotcurshape} dispenses with
dividing the galaxy into rings and instead fits the whole velocity
field at once.  In addition to calculating the best-fit values for the
PA, inclination, systemic velocity, and center, \rotcurshape\ also
assumes a functional form for the rotation curve/density profile
(e.g., power law, NFW, pseudoisothermal, etc.) and solves for the free
parameters of that function.  One advantage of this approach is that
near the center of the galaxy, where the velocities may be changing
rapidly with radius, all of the data points are not artificially
placed at the same radius (as was necessary with \rotcur, where every
point with $r < 8\arcsec$ was in the same ring).  Another is that the
kinematic parameters of the galaxy and the parameters of the fitting
function are determined in a single step.  This makes it
straightforward to measure the global agreement between the model and
the data.  For a power law rotation curve, the results from
\rotcurshape\ are nearly identical to the ones we derive by running
\rotcur\ or \ringfit\ and then fitting a power law to the resulting
rotation velocities.

\section{NFW AND POWER LAW DENSITY PROFILES}
\label{densityprofs}

\citet*{nfw96} showed that CDM halos have density profiles of the form

\begin{equation}
\frac{\rho(r)}{\rho_{crit}} = \frac {\delta_c}{(r/r_s)(1+r/r_s)^2} \ ,
\label{nfw}
\end{equation}

\noindent where $\rho_{crit}=3H^2/8\pi G\sim10^{-29}$ \gpercmcu\ is
the critical density, $\delta_c$ is the halo overdensity, and $r_s$ is
the scale radius (simulations suggest $r_s \sim 2.5$ kpc for a galaxy
the size of NGC~2976).  For $r \ll r_{s}$, Equation \ref{nfw} clearly
reduces to $\rho \propto r^{-1}$.  The commonly-discussed
concentration parameter $c$ is the ratio of the virial radius of the
halo ($r_{200}$, the radius enclosing a mean density of 200 times the
background density) to the scale radius.  In the simulations analyzed
by NFW the concentration parameter varied from $\sim7$ for galaxy
clusters up to $\sim16$ for large galaxies.  Later studies at lower
masses found a median concentration of $c = 20.5$ for $3 \times
10^{10} M_{\odot}$ halos \citep{bullock01}.  Other numerical
simulations have resulted in slightly different profile shapes.  For
example, \citet{moore99b} argue that CDM halos exhibit steeper central
cusps when simulated at higher resolution; their best-fitting
functional form is similar to that of NFW, except that both terms in
the denominator of the right hand side of Equation \ref{nfw} are
raised to the 1.5 power, resulting in a $\rho \propto r^{-1.5}$
central density profile.  Most subsequent studies in the literature
have found central slopes that are bracketed by the NFW and Moore
profiles \citep[e.g.,][]{js00,g00,kkbp01,power03}.  It is noteworthy
that no set of simulations has found central density profiles that are
shallower than $\rho \propto r^{-1}$, although \citet{tn01} presented
analytical arguments for a $\rho \propto r^{-0.75}$ central slope.

Since we are mostly interested in power law fits to the rotation
curve, we also note that for a spherical mass distribution, a density
profile $\rho = \rho_{0} (r/r_{0})^{-\alpha}$ implies that

\begin{equation}
v_{rot} = \sqrt{\frac{4\pi G\rho_{0} r_{0}^{2}}{3-\alpha}} 
\left( \frac{r}{r_{0}} \right) ^{(2-\alpha)/2} \ ,
\label{alphaeq}
\end{equation}

\noindent and correspondingly, a rotation curve that can be fit by a
power law $v_{rot} = v_{0} (r/r_{0})^{\beta}$ yields a density profile

\begin{equation}
\rho = \frac{(2\beta + 1)v_{0}^{2}}{4\pi Gr_{0}^{2}} 
\left( \frac{r}{r_{0}} \right) ^{2\beta-2} \ .
\label{betaeq}
\end{equation}

\noindent A galaxy with a constant density halo thus has a linear
($v_{rot} \propto r$) rotation curve, while the rotation curve
associated with an NFW $\rho \propto r^{-1}$ central density profile
is $v_{rot} \propto r^{1/2}$.

\begin{deluxetable}{l c c c c c c}
\tablenum{1}
\tablewidth{0pt}
\tablecolumns{6}
\tablecaption{NGC~2976 Surface Brightness Profiles}
\tablehead{
\colhead{Filter} &
\colhead{Integrated} & \colhead{Central Surface Brightness\tablenotemark{b}} &
\colhead{Inner Disk} & \colhead{Outer Disk} &
\colhead{$\mu_{\mbox{sky}}$} \\
\colhead{} & \colhead{Magnitude\tablenotemark{a}} &
\colhead{$\mu_{0}$ [\surfb]} &
\colhead{Scale Length\tablenotemark{b}\phm{00}[\arcsec]} &
\colhead{Scale Length\tablenotemark{c}\phm{00}[\arcsec]} & \colhead{[\surfb]} }

\startdata
B & 10.71       & $21.31 \pm 0.01$  & $79.3 \pm 1.5$  & $34.1 \pm 0.4$  & 22.10 \\
V & 10.14       & $20.69 \pm 0.01$  & $72.8 \pm 0.8$  & $33.6 \pm 0.2$  & 21.28 \\
R & \phm{1}9.66 & $20.21 \pm 0.01$  & $71.2 \pm 0.7$  & $34.4 \pm 0.2$  & 20.71 \\
I & \phm{1}9.19 & $19.73 \pm 0.01$  & $69.8 \pm 0.7$  & $33.1 \pm 0.1$  & 19.68 \\
J & \phm{1}8.29 & $18.88 \pm 0.03$  & $71.8 \pm 4.0$  & $34.3 \pm 2.9$  & 16.01 \\
H & \phm{1}7.71 & $18.24 \pm 0.04$  & $70.2 \pm 5.5$  & $31.1 \pm 3.9$  & 13.98 \\
K$_{\mbox{s}}$ & \phm{1}7.48 & $18.03 \pm 0.06$  & $69.6 \pm 7.2$  & $31.5 \pm 5.0$  & 13.47

\enddata
\label{disktab}

\tablenotetext{a}{These magnitudes are measured within an elliptical
aperture with a semimajor axis of 172\arcsec\ on our Lowell and 2MASS
images.  The galaxy does extend to somewhat larger radii on the Keck
images, so we have certainly underestimated the flux here.  The Keck
data suggest that the Lowell magnitudes should be made $\sim4$~\%
brighter, although if the galaxy is more extended even than those
images reveal, the true correction could be slightly larger.}
\tablenotetext{b}{Central surface brightnesses and inner scale lengths
were calculated from the light distribution between 10\arcsec\ and
70\arcsec.}  
\tablenotetext{c}{Outer scale lengths were calculated from the light
distribution outside 100\arcsec\ for B, V, and R (where there was a
visible transition region between the inner and outer disks), and
outside 70\arcsec\ for the near-infrared bands (where there was no
transition region).}

\tablecomments{We have applied Galactic extinction corrections to
these data.  Internal extinction corrections have \emph{not} been
applied, but our adopted values for the internal extinction are given
in the text (\S \ref{sbprofsec}) if the reader wishes to use them.}

\end{deluxetable}

\begin{deluxetable}{c c c c c c c c}
\tablenum{2}
\tablewidth{0pt}
\tablecolumns{8}
\tablecaption{Rotation Curve Data}
\tablehead{ \colhead{Radius\tablenotemark{a}} & \colhead{$v_{rot}$\tablenotemark{b}} &
\colhead{$v_{rad}$\tablenotemark{b,c}} & \colhead{$v_{sys}$\tablenotemark{b,c}} &
\colhead{$v_{*,rot}$\tablenotemark{d}} & \colhead{$v_{\mbox{\tiny HI},rot}$\tablenotemark{e}} &
\colhead{$v_{\mbox{\tiny CO},rot}$\tablenotemark{f}} & \colhead{$\Delta v_{drift}$\tablenotemark{g}} \\
\colhead{[\arcsec]} & \colhead{[\kms]} & \colhead{[\kms]} &
\colhead{[\kms]} & \colhead{[\kms]} & \colhead{[\kms]} &
\colhead{[\kms]} & \colhead{[\kms]} }

\startdata
6.2  & \phn$6.8 \pm 0.4 \pm 3.6$  & \phs\phn$3.8 \pm 0.2 \pm 3.3$   & $-0.1 \pm 0.1 \pm 3.1$    & $20.6 \pm 0.5$ & 0.9  & 2.4 & \phs$4.7$ \\
10.0 & \phn$9.5 \pm 0.3 \pm 4.3$  & \phs\phn$8.4 \pm 0.1 \pm 3.8$   & $-2.1 \pm 0.1 \pm 4.0$    & $23.8 \pm 0.6$ & 1.5  & 4.1 & \phs$4.4$ \\
14.0 & $14.0 \pm 0.2 \pm 3.1$  & \phs$12.4 \pm 0.1 \pm 2.8$  & $-3.8 \pm 0.1 \pm 2.7$    & $27.9 \pm 0.7$ & 2.1  & 5.0 & \phs$3.1$ \\
18.1 & $19.8 \pm 0.1 \pm 5.5$  & \phs$14.2 \pm 0.1 \pm 2.4$  & $-2.8 \pm 0.1 \pm 1.6$    & $33.4 \pm 0.8$ & 2.5  & 5.0 & \phs$1.9$ \\
22.1 & $26.1 \pm 0.1 \pm 3.9$  & \phs$15.7 \pm 0.1 \pm 2.7$  & $-1.1 \pm 0.1 \pm 1.1$    & $38.1 \pm 0.8$ & 2.8  & 5.1 & \phs$1.0$ \\
26.0 & $28.7 \pm 0.1 \pm 3.5$  & \phs$15.4 \pm 0.1 \pm 2.7$  & $-1.0 \pm 0.1 \pm 1.0$    & $41.2 \pm 0.7$ & 3.4  & 5.0 & \phs$0.4$ \\
30.0 & $28.7 \pm 0.1 \pm 2.7$  & \phs$13.0 \pm 0.1 \pm 2.6$  & $-1.3 \pm 0.1 \pm 1.2$    & $46.8 \pm 0.7$ & 3.8  & 4.9 & $-0.1$ \\
34.0 & $31.7 \pm 0.1 \pm 2.5$  & \phs$10.6 \pm 0.1 \pm 2.4$  & $-0.5 \pm 0.1 \pm 1.7$    & $50.0 \pm 0.6$ & 4.2  & 4.8 & $-0.5$ \\
38.0 & $35.5 \pm 0.1 \pm 2.3$  & \phs\phn$9.5 \pm 0.1 \pm 2.3$   & \phs$0.1 \pm 0.1 \pm 1.5$ & $52.9 \pm 0.5$ & 5.0  & 4.7 & $-0.8$ \\
42.0 & $39.4 \pm 0.1 \pm 2.1$  & \phs\phn$8.0 \pm 0.1 \pm 2.7$   & \phs$0.2 \pm 0.1 \pm 1.3$ & $57.0 \pm 0.4$ & 5.9  & 5.1 & $-1.1$ \\
46.0 & $42.9 \pm 0.1 \pm 2.3$  & \phs\phn$7.1 \pm 0.1 \pm 3.1$   & $-0.1 \pm 0.1 \pm 1.3$    & $60.5 \pm 0.2$ & 6.4  & 4.9 & $-1.3$ \\
50.0 & $46.0 \pm 0.1 \pm 2.6$  & \phs\phn$6.7 \pm 0.1 \pm 3.4$   & $-0.8 \pm 0.1 \pm 1.5$    & $64.0 \pm 0.1$ & 7.2  & 4.6 & $-1.5$ \\
54.0 & $49.1 \pm 0.1 \pm 3.1$  & \phs\phn$5.5 \pm 0.1 \pm 3.5$   & $-1.8 \pm 0.1 \pm 1.9$    & $65.6 \pm 0.1$ & 7.9  & 5.2 & $-1.5$ \\
58.1 & $49.2 \pm 0.1 \pm 3.2$  & \phs\phn$4.6 \pm 0.1 \pm 3.6$   & $-3.0 \pm 0.1 \pm 2.2$    & $67.9 \pm 0.3$ & 8.6  & 5.0 & $-1.5$ \\
62.1 & $51.4 \pm 0.1 \pm 3.2$  & \phs\phn$5.3 \pm 0.1 \pm 3.1$   & $-3.4 \pm 0.1 \pm 1.8$    & $71.9 \pm 0.4$ & 9.6  & 5.1 & $-1.3$ \\
66.0 & $57.2 \pm 0.1 \pm 3.1$  & \phs\phn$6.0 \pm 0.1 \pm 3.0$   & $-3.8 \pm 0.1 \pm 1.7$    & $73.2 \pm 0.6$ & 11.1 & 5.9 & $-0.8$ \\
69.9 & $63.8 \pm 0.1 \pm 2.9$  & \phs\phn$5.4 \pm 0.1 \pm 3.6$   & $-3.0 \pm 0.1 \pm 1.5$    & $76.3 \pm 0.8$ & 13.2 & 5.9 & $-0.3$ \\
73.9 & $67.8 \pm 0.1 \pm 2.8$  & \phs\phn$3.7 \pm 0.1 \pm 4.7$   & $-1.7 \pm 0.1 \pm 1.4$    & $80.7 \pm 1.0$ & 15.0 & 6.0 & \phs$0.4$ \\
77.9 & $69.8 \pm 0.1 \pm 3.3$  & \phs\phn$2.1 \pm 0.1 \pm 5.8$   & \phs$0.0 \pm 0.1 \pm 1.7$ & $83.5 \pm 1.2$ & 16.7 & 6.2 & \phs$1.1$ \\
81.9 & $71.7 \pm 0.1 \pm 3.9$  & \phs\phn$1.5 \pm 0.1 \pm 6.6$   & \phs$2.3 \pm 0.1 \pm 2.2$ & $84.9 \pm 1.4$ & 18.5 & 6.4 & \phs$1.9$ \\
85.9 & $74.1 \pm 0.1 \pm 4.4$  & \phs\phn$0.0 \pm 0.2 \pm 7.9$   & \phs$3.5 \pm 0.1 \pm 2.7$ & $84.5 \pm 1.6$ & 20.2 & 6.5 & \phs$2.7$ \\
89.9 & $76.7 \pm 0.1 \pm 4.8$  & \phn$-3.8 \pm 0.3 \pm 9.1$  & \phs$2.7 \pm 0.1 \pm 2.8$ & $83.5 \pm 1.8$ & $22.1$ & 6.7 & \phs$3.4$ \\
94.0 & $79.9 \pm 0.2 \pm 4.4$  & $-8.3 \pm 0.4 \pm 12.4$ & \phs$0.1 \pm 0.1 \pm 2.8$ & $85.4 \pm 2.0$ & $23.6$ & 6.9 & \phs$4.1$ \\
100.6 & $83.6 \pm 0.6 \pm 3.3$ & 0                           & 0                         & $84.5 \pm 2.3$ & $25.7$ & 7.1 & \phs$5.0$ \\
111.1 & $83.9 \pm 0.6 \pm 3.9$ & 0                           & 0                         & $81.2 \pm 2.8$ & $25.1$ & 7.5 & \phs$5.7$ \\
120.7 & $88.7 \pm 0.6 \pm 4.4$ & 0                           & 0                         & $80.8 \pm 3.3$ & $25.3$ & 7.8 & \phs$4.6$ \\
131.2 & $85.3 \pm 0.8 \pm 5.3$ & 0                           & 0                         & $77.6 \pm 3.8$ & $25.4$ & 8.1 & \phs$3.2$ 

\enddata
\label{data}

\tablenotetext{a}{To convert to pc, multiply by 16.7.}
\tablenotetext{b}{Fitted velocities are given as value $\pm$
statistical error $\pm$ systematic error.}  
\tablenotetext{c}{Radial velocities and systemic velocities were fixed
at zero for the outer four rings, where a lack of velocity field
information away from the major axis limited our ability to constrain
them.}
\tablenotetext{d}{Stellar velocities are given for the case of \mslk
$= 1.0 \mlk$.  To get the stellar velocities for a different stellar
mass-to-light ratio, multiply the tabulated values by $\sqrt{\mslk}$.
The listed uncertainties include only statistical errors.}
\tablenotetext{e}{The uncertainties on the \hi\ rotation velocities
are not known because we do not have access to the original data, but
are probably about 10~-~20~\%.}  \tablenotetext{f}{The uncertainties
on the CO rotation velocities are quite high because the CO-\htwo\
conversion factor is not known accurately.  Since the CO rotation
velocities are so small, this uncertainty is unimportant.}
\tablenotetext{g}{This column gives the asymmetric drift correction to
the rotation curve (see \S \ref{asymdrift}).  To correct for
asymmetric drift, add the values in this column to the observed
rotation velocities in column 2.}

\end{deluxetable}

\begin{deluxetable}{c c c c c}
\tablenum{3}
\tablewidth{0pt}
\tablecolumns{5}
\tablecaption{Stellar Mass-to-Light Ratio Predictions}
\tablehead{
\colhead{Color} & \colhead{Mean Inner} &
\colhead{Predicted \mslk\tablenotemark{c}} & \colhead{Predicted \mslr} &
\colhead{Mean Outer} \\
\colhead{} & \colhead{Disk Color\tablenotemark{a,b}} &
\colhead{[\mlk]} & \colhead{[\mlr]} & \colhead{Disk Color\tablenotemark{b,d}}}

\startdata
$B-V$ &  0.53  &  0.45  &  0.97  & 0.60 \\
$B-R$ &  0.98  &  0.46  &  1.03  & 1.10 \\
$V-I$ &  0.87  &  0.47  &  1.04  & 0.93 \\
$V-J$ &  1.72  &  0.49  &  1.11  & 1.85 \\
$V-H$ &  2.31  &  0.50  &  1.16  & 2.33 \\
$V-K$ &  2.48  &  0.49  &  1.13  & 2.53

\enddata

\tablenotetext{a}{Calculated for $10\arcsec \le r \le 70\arcsec$.}
\tablenotetext{b}{Note that these colors have been corrected for
Galactic extinction and internal extinction.  The Galactic extinction
is taken from \citet*{sfd98} and the internal extinction corrections
are given in the text.}
\tablenotetext{c}{We use the predictions for the formation epoch with
bursts of star formation model, assuming a scaled Salpeter initial
mass function, as described in \citet{bdj01}.}
\tablenotetext{d}{Calculated for $100\arcsec \le r \le 172\arcsec$.}
\label{mltable}

\end{deluxetable}

\begin{deluxetable}{l c c c c}
\tablenum{4}
\tablewidth{0pt}
\tablecolumns{5}
\tablecaption{Central Positions for NGC~2552}
\tablehead{
\colhead{Method of} & \colhead{$\alpha$ (J2000.0)} & \colhead{$\delta$
(J2000.0)} & \colhead{Distance From}  & \colhead{Reference}\\
\colhead{Determining Center} & \colhead{} & \colhead{} & \colhead{Nucleus} [\arcsec] & \colhead{}}

\startdata
nucleus\tablenotemark{a}         & 08\hr19\min20.4\sec & 50\degr00\arcmin33\arcsec & \phm{1}0.0 & 1 \\
outer isophotes\tablenotemark{b} & 08\hr19\min20.4\sec & 50\degr00\arcmin36\arcsec & \phm{1}2.7 & 2 \\
outer isophotes\tablenotemark{c} & 08\hr19\min19.7\sec & 50\degr00\arcmin32\arcsec & \phm{1}6.8 & 3 \\
NED                              & 08\hr19\min20.1\sec & 50\degr00\arcmin25\arcsec & \phm{1}8.5 & 4 \\
SIMBAD                           & 08\hr19\min19.6\sec & 50\degr00\arcmin28\arcsec & \phm{1}9.2 & 5

\enddata

\tablenotetext{a}{This position is also the center of the inner
isophotes.}
\tablenotetext{b}{Measured by \citet{swaters99} from their photometry.}
\tablenotetext{c}{Measured by \citet{sb02} from their photometry.}

\tablerefs{1, \citet*{bsvdm03}; 2, \citet{vdbs01}; 3,
\citet{swaters02}; 4, \citet{uzc}; 5, \citet*{ugc}.}

\label{NGC2552table}
\end{deluxetable}

\end{document}